\documentclass[12pt]{article}
\begin{document}  

% MANIFOLDS

\newcommand{\yman}{\ensuremath{\cal Y}}
\newcommand{\xman}{\ensuremath{\cal X}}
\newcommand{\tman}{\ensuremath{\cal T}}
\newcommand{\zman}{\ensuremath{\cal Z}}
\newcommand{\fman}{\ensuremath{\cal F}}
\newcommand{\mman}{\ensuremath{\cal M}}

% SPTM FIELDS

\newcommand{\F}{\ensuremath{\sf \Phi}}
\newcommand{\G}{\ensuremath{\sf G}}
\newcommand{\T}{\ensuremath{\sf T}}
\newcommand{\X}{\ensuremath{\sf X}}
\newcommand{\Z}{\ensuremath{\sf Z}}
\newcommand{\E}{\ensuremath{\sf E}}
\newcommand{\PF}{\ensuremath{\sf {{\Pi}_{\F}}}}
\newcommand{\PG}{\ensuremath{\sf {{\Pi}_{\G}}}}
\newcommand{\PT}{\ensuremath{\sf {{\Pi}_{\T}}}}
\newcommand{\PX}{\ensuremath{\sf {{\Pi}_{\X}}}}
\newcommand{\PZ}{\ensuremath{\sf {{\Pi}_{\Z}}}}

% COORDINATE SPTM FIELDS

\newcommand{\FA}{\ensuremath{{\sf \Phi}^{A}}}
\newcommand{\FB}{\ensuremath{{\sf \Phi}^{B}}}
\newcommand{\FC}{\ensuremath{{\sf \Phi}^{C}}}
\newcommand{\Gab}{\ensuremath{{\sf G}_{\alpha\beta}}}
\newcommand{\Gmn}{\ensuremath{{\sf G}_{\mu\nu}}}
\newcommand{\iGab}{\ensuremath{{\sf G}^{\alpha\beta}}}
\newcommand{\iGmn}{\ensuremath{{\sf G}^{\mu\nu}}}
\newcommand{\Gam}{\ensuremath{{\sf G}_{\alpha\mu}}}
\newcommand{\Gan}{\ensuremath{{\sf G}_{\alpha\nu}}}
\newcommand{\Gbm}{\ensuremath{{\sf G}_{\beta\mu}}}
\newcommand{\Gbn}{\ensuremath{{\sf G}_{\beta\nu}}}
\newcommand{\iGam}{\ensuremath{{\sf G}^{\alpha\mu}}}
\newcommand{\iGan}{\ensuremath{{\sf G}^{\alpha\nu}}}
\newcommand{\iGbm}{\ensuremath{{\sf G}^{\beta\mu}}}
\newcommand{\iGbn}{\ensuremath{{\sf G}^{\beta\nu}}}
\newcommand{\Tt}{\ensuremath{  {\sf T}^{0} } }
\newcommand{\Xx}{\ensuremath{  {\sf X}^{1} } }
\newcommand{\Za}{\ensuremath{{{\sf Z}^{a}}}}
\newcommand{\Zb}{\ensuremath{{{\sf Z}^{b}}}}

\newcommand{\PA}{\ensuremath{{\sf \Pi}_{A}}}
\newcommand{\PB}{\ensuremath{{\sf \Pi}_{B}}}
\newcommand{\PC}{\ensuremath{{\sf \Pi}_{C}}}
\newcommand{\Pab}{\ensuremath{{\sf \Pi}^{\alpha\beta}}}
\newcommand{\Pmn}{\ensuremath{{\sf \Pi}^{\mu\nu}}}
\newcommand{\Pgd}{\ensuremath{{\sf \Pi}^{\gamma\delta}}}
\newcommand{\Pt}{\ensuremath{{\sf \Pi}_{0}}}
\newcommand{\Px}{\ensuremath{{\sf \Pi}_{1}}}
\newcommand{\Pa}{\ensuremath{{\sf \Pi}_{a}}}
\newcommand{\Pb}{\ensuremath{{\sf \Pi}_{b}}}

% PULLBACK FIELDS

\newcommand{\f}{\ensuremath{\Phi}}
\newcommand{\g}{\ensuremath{G}}
\newcommand{\pf}{\ensuremath{{{\Pi}_{\f}}}}
\newcommand{\pg}{\ensuremath{{{\Pi}_{\g}}}}

% COORDINATE PULLBACK FIELDS

\newcommand{\ya}{\ensuremath{Y^{\alpha}}}
\newcommand{\yb}{\ensuremath{Y^{\beta}}}

\newcommand{\gtt}{\ensuremath{G_{00}}}
\newcommand{\igtt}{\ensuremath{G^{00}}}
\newcommand{\gtx}{\ensuremath{G_{01}}}
\newcommand{\igtx}{\ensuremath{G^{01}}}
\newcommand{\gxx}{\ensuremath{G_{11}}}
\newcommand{\igxx}{\ensuremath{G^{11}}}

\newcommand{\fa}{\ensuremath{\Phi}^A}
\newcommand{\fb}{\ensuremath{\Phi}^B}
\newcommand{\fc}{\ensuremath{\Phi}^C}
\newcommand{\gab}{\ensuremath{{G}_{ab}}}
\newcommand{\gmn}{\ensuremath{{G}_{mn}}}
\newcommand{\igab}{\ensuremath{{G}^{ab}}}
\newcommand{\igmn}{\ensuremath{{G}^{mn}}}

\newcommand{\pa}{\ensuremath{{\Pi}_A}}
\newcommand{\pb}{\ensuremath{{\Pi}_B}}
\newcommand{\pc}{\ensuremath{{\Pi}_C}}
\newcommand{\pab}{\ensuremath{{\Pi}^{ab}}}
\newcommand{\pmn}{\ensuremath{{\Pi}^{mn}}}

\newcommand{\ptt}{\ensuremath{{\Pi}^{00}}}
\newcommand{\ptx}{\ensuremath{{\Pi}^{01}}}
\newcommand{\pxx}{\ensuremath{{\Pi}^{11}}}

\newcommand{\qxx}{\ensuremath{Q_{11}}}
\newcommand{\iqxx}{\ensuremath{H^{11}}}

\newcommand{\lie}{\ensuremath{{\cal L}_{({\sf E})}}}

\newcommand{\Ea}{\ensuremath{{\sf E}^{\alpha}}}
\newcommand{\Eb}{\ensuremath{{\sf E}^{\beta}}}
\newcommand{\Em}{\ensuremath{{\sf E}^{\mu}}}
\newcommand{\En}{\ensuremath{{\sf E}^{\nu}}}

\newcommand{\Qab}{\ensuremath{{\sf Q}^{\alpha\beta}}}
\newcommand{\Qbc}{\ensuremath{{\sf Q}^{\beta\gamma}}}
\newcommand{\Qam}{\ensuremath{{\sf Q}^{\alpha\mu}}}
\newcommand{\Qan}{\ensuremath{{\sf Q}^{\alpha\nu}}}

\newcommand{\N}{\ensuremath{N_0}}
\newcommand{\M}{\ensuremath{M^{1}_{0}}}
\newcommand{\h}{\ensuremath{H_{11}}}
\newcommand{\ih}{\ensuremath{H^{11}}}

%%%%%%%%%%%%%%%%%%%%%%%%%%%%%%%%%%%%%%%%%%%%%%%%%%%%%%%%

\titlepage

\title{Diffeomorphisms as Symplectomorphisms\\ in History 
Phase Space:\\ Bosonic String Model}   

\author{I. Kouletsis\thanks{Present address: Institut 
f\"{u}r theoretische Physik, Universit\"{a}t Bern, CH-3012 
Bern, Switzerland}$\;$  and K. V. Kucha\v{r}\\{}\\
Department of Physics\\University of Utah\\Salt Lake City\\
Utah 84109} 

\maketitle

%%%%%%%%%%%%%%%%%%%%%%%%%%%%%%%%%%%%%%%%%%%%%%%%%%%%%%%%

\renewcommand{\theequation}{\thesection.\arabic{equation}}
\let\ssection=\section
\renewcommand{\section}{\setcounter{equation}{0}\ssection}

\begin{abstract}

The structure of the history phase space $\cal G$ of a
covariant field system and its history group (in the sense
of Isham and Linden) is analyzed on an example of a bosonic
string. The history space $\cal G$ includes the time map
$\sf T$ from the spacetime manifold (the two-sheet) $\cal Y$
to a one-dimensional time manifold $\cal T$ as one of its
configuration variables. A canonical history action is
posited on $\cal G$ such that its restriction to the
configuration history space yields the familiar Polyakov
action. The standard Dirac-ADM action is shown to be
identical with the canonical history action, the only
difference being that the underlying action is expressed in
two different coordinate charts on $\cal G$. The canonical
history action encompasses all individual Dirac-ADM actions
corresponding to different choices $\sf T$ of foliating
$\cal Y$. The history Poisson brackets of spacetime fields on
$\cal G$ induce the ordinary Poisson brackets of spatial
fields in the instantaneous phase space ${\cal G}_{0}$ of
the Dirac-ADM formalism. The canonical history action is
manifestly invariant both under spacetime diffeomorphisms
Diff$\cal Y$ and temporal diffeomorphisms Diff$\cal T$. Both
of these diffeomorphisms are explicitly represented by
symplectomorphisms on the history phase space $\cal G$. The
resulting classical history phase space formalism is offered
as a starting point for projection operator quantization and
consistent histories interpretation of the bosonic string
model.

\end{abstract}

\section{Introduction}

\subsection{Framework}

Conventional interpretation of quantum theory conceives a
quantum system described by the density operator $\widehat
\rho$ on a Hilbert space $\cal H$, and classical apparatuses
designed to perform instantaneous measurements of different
properties of the system characterized by projection
operators $\widehat \alpha$ on $\cal H$. The probability
that the system which is in the state $\widehat \rho$ will
be found to have the property $\widehat \alpha$ in a single
instantaneous measurement is given by von Neumann's formula 
\cite{vn}
\begin{equation}
{\rm Prob} ({\widehat \alpha}, {\widehat \rho}\,) = 
{\rm Tr}_{\cal H}\, ({\widehat \alpha}\, {\widehat \rho}\, 
{\widehat \alpha}^\dagger) \, .           \label{eq:vneu}
\end{equation}

In between consecutive measurements, the state of the system
is supposed to be changed by the evolution operator
\begin{equation}
{\widehat U} (t, t') = {\rm e}^{-i {\widehat H} (t-t')}
                              \label{eq:evol}
\end{equation}
generated by the Hamiltonian $\widehat H$. A {\it
homogeneous history} of the system is a time-ordered
sequence of properties
\begin{equation}
\alpha = \big ( {\widehat \alpha}_{t_1},\, {\widehat
\alpha}_{t_2},\,...\, , {\widehat \alpha}_{t_n}\big )\,, \;\;
\; t_1<t_2<\cdot \cdot \cdot <t_n \,.
\end{equation}
These can be strung by the evolution operator $\widehat U$
into a single operator
\begin{equation}
{\widehat C}_{\alpha} = {\widehat U} (t_0, t_n)\, {\widehat
\alpha}_{t_n} {\widehat U} (t_n, t_{n-1})\;\cdot \cdot
\cdot\; {\widehat U}\, (t_2, t_1)\, {\widehat \alpha}_{t_1}
{\widehat U} (t_1, t_{0}) \label{eq:clop}
\end{equation}
on $\cal H$, called the {\it class operator}. The state of
the system $\widehat \rho \,$ is prescribed at an initial
instant $t_0<t_1\,$. The probability that the appropriate
sequence of measurements will detect the history $\alpha$ is
then neatly summarized by the Groenewold-Wigner formula 
\cite{gro}, \cite{wig}
\begin{equation}
{\rm Prob} (\alpha, {\widehat \rho}\,) = {\rm Tr}_{\cal H}
({\widehat C}_{\alpha}\, {\widehat \rho}\, {\widehat
C}_{\alpha}^{\dagger})\,.    \label{eq:grwig}
\end{equation}

Consistent histories interpretation of quantum theory,
proposed and developed by Griffiths \cite{g}, Omn\`{e}s
\cite{o}-\cite{o4}, and Gell-Mann and Hartle \cite{gh}-\cite{gh5} asserts that,
under appropriate conditions, the probability assignement
(\ref{eq:grwig}) is still meaningful for a closed quantum
system without any need to introduce external classical
apparatuses. Conditions ensuring that a set of histories is
consistent and can be assigned probabilities are based on
studying the components of the {\it decoherence functional}
\begin{equation}
d_{({\hat H}, {\hat \rho})} (\alpha, \beta) =
{\rm Tr}_{\cal H}(
{\widehat C}_{\alpha}\, {\widehat \rho}\, 
{\widehat C}_{\beta}^{\dagger})   \label{eq:defun}
\end{equation}
between any two histories $\alpha$ and $\beta$ of the
set. The details of these conditions do not concern us in
this paper.

In spite of their similarity, the Groenewald-Wigner formula
(\ref{eq:grwig}) differs from the von Neumann formula
(\ref{eq:vneu}) in two important respects. First, unlike
$\widehat \alpha$, ${\widehat C}_{\alpha}$ is not in general
a projection operator and hence not directly amenable to the
standard operations `or' and `not' of quantum logic.
Second, the von Neumann formula is purely kinematical, while
in the Groenewold-Wigner formula the kinematical aspects
(the histories $\alpha$ which are being studied) and the
dynamical aspects (the evolution $\widehat U$ of the state)
are intertwined in the class operator (\ref{eq:clop}). Both of
these inconvenient features have been removed in the
projection operator reformulation of the Gell-Mann and
Hartle scheme by Isham \cite{isham} and Isham and Linden \cite{il}.

The key idea of this reformulation is to 
represent homogeneous histories by genuine projection operators
\begin{equation}
{\widehat \alpha} = {\widehat \alpha}_{t_1} \otimes
 {\widehat \alpha}_{t_2} \otimes \, \cdot \cdot \cdot \,
 \otimes {\widehat \alpha}_{t_n} \label{eq:hispro}
\end{equation}
on the tensor product ${\cal N}_n = {\cal H}_{t_1} \otimes
{\cal H}_{t_2} \otimes \, ... \, \otimes {\cal H}_{t_n}$ of
time-labeled replicas of the original Hilbert space $\cal
H$. This allows one to combine two (disjoint) homogeneous
projectors $\widehat \alpha$ and $\widehat \beta$ into a new
projector ${\widehat \gamma} ={\widehat \alpha}\, \oplus \,
{\widehat \beta}$ which is no longer necessarily
homogeneous, i.e., expressible as a tensor product
(\ref{eq:hispro}) of projectors on $\cal H$. The projector
$\widehat \gamma$ represents the history proposition ``the
history $\alpha$ is realized or the history $\beta$ is
realized'' and its inhomogeneous character encodes the
concept of `temporal entanglement' of instantaneous
properties. Isham and Linden also succeeded to separate
kinematics from dynamics by casting the Gell-Mann and Hartle
decoherence functional (\ref{eq:defun}) in the form in which
histories and dynamics enter through two separate operators
on the tensor product Hilbert space.

What concerns us most, however, is the third achievement of
the projector operator formalism, namely, its extension to
{\it continuous histories}. This was reached by constructing
a continuous tensor product ${\cal N} = {\otimes}_t \, {\cal
H}_t$ of copies ${\cal H}_t$ of the Hilbert space $\cal
H$. In the particular construction used by Isham and Linden,
${\cal N}$ is equivalent to the temporal bosonic Fock space
of the instantaneous Hilbert space of the
system. \cite{il2} This Fock space carries a representation
of certain kinematical history operators, like continuous
products of projectors into coherent states, or of more
directly interpretable continuous time averages of
projectors into ranges of fundamental canonical
variables. It was also shown \cite{ilss} to be able to carry
the dynamics of simple systems like a harmonic
oscillator and a certain wider class of models, by
representing propositions about time-averaged values of the
Hamiltonian.

The program of constructing a Hilbert space ${\cal N}$
appropriate for carrying both the history propositions and
dynamics of a given system can be posed in a
similar way as the problem of constructing the instantaneous
Hilbert space $\cal H$ of the system. A convenient
formulation of the latter problem is in terms of
representations of the Weyl-Heisenberg group. 

Consider a finite-dimensional classical system moving in the
phase space $T^{*} {\cal Q}$ which is the cotangent bundle
of a configuration space $\cal Q$ and possesses a natural
coordinate chart $q^a, p_a \,$. The Weyl-Heisenberg group of
the system is generated by the elements $q^a$, $p_a$ and $1$
of the Lie algebra
\begin{equation}
\big \{ q^a, q^b \big \} = 0\,, \; \big \{ p_a , p_b \big \} =0
\, , \; \big \{ q^a , p_b \big \} = \delta ^a_b \, .
\label{eq:w-al}
\end{equation}
Canonical quantization of the system consists in the task of
finding a Hilbert space $ {\cal H}$ which carries an
irreducible unitary representation of the Weyl-Heisenberg
group. The Stone and von Neumann theorem shows that the
solution of this task is essentially unique. The
construction of $\cal H$ automatically yields a privileged
set of projection operators on $\cal H$, namely those into
different ranges of the spectra of the fundamental variables
${\widehat q}^a$ and ${\widehat p}_a \,$. These projection
operators can be considered as the fundamental properties of
the quantum system. To represent the Hamiltonian $H$ of
standard dynamical systems by a self-adjoint operator
$\widehat H$ on $\cal H$ does not present in principle any
difficulty.

Following the same line of reasoning, Isham and Linden
turned the task of finding the history Hilbert space into a
problem of constructing a Hilbert space ${\cal N}$ that
carries an irreducible unitary representation of the {\it
history group} whose generators $q^a(t),\, p_a(t)$ and $1$
obey the Lie algebra
\begin{equation}
\Big \{ \!\!\! \Big \{ q^a(t), q^b(t') \, \Big \}\!\! \! \Big \} = 0\,,\;
\Big \{\!\! \! \Big \{ p_a(t), p_b(t') \, \Big \}\!\! \! \Big \} = 0\,,\;
\Big \{ \!\!\! \Big \{ q^a(t), p_b(t') \, \Big \} \!\!\! \Big \} = 
\delta^a_b \, \delta(t,t')       \label{eq:hisgr}
\end{equation}
reminiscent of a quantum field theory in one dimension which
is spanned by the time variable. Because the Stone and von
Neumann theorem does not hold for infinitely-dimensional
systems, there may exist many unitarily non-equivalent
representations of the history group (\ref{eq:hisgr}). One
expects the appropriate space ${\cal N}$ to be selected by the
requirement that it also carry a representation of the
time-averaged values of the Hamiltonian. The results
reported by Isham {\it et al.} \cite{ilss} show that the
ordinary bosonic Fock space is able to carry both the
fundamental history propositions and the dynamics of certain
simple finite-dimensional systems. Further investigations on the
continuous histories of such systems have been carried out by 
Savvidou \cite{s}, \cite{s2}. 

%%%%%%%%%%%%%%%%%%%%%%%%%%%%%%%%%%%%%%%%%%%%%%%%%%%%%%%%%%%%%

\subsection{Goals and Results}

We are pursuing the problem of how to handle covariant
dynamical systems (a relativistic particle, a string,
general relativity, or parametrized particle and field
theories) in the consistent histories framework. The
Gell-Mann and Hartle formalism addressed this question
and reached many valuable insights. Covariant systems were
handled by means of path integrals in the configuration
space with the implication that history propositions 
were restricted to the configuration variables while the
momenta were considered as derivative quantities emerging in
particular arrangements of the system. The projection
operator formalism of Isham and Linden treats the
fundamental phase space variables on an equal footing. Its
application to parametrized systems was considered by
Anastopoulos and Savvidou \cite{sa}, and to general relativity 
by Savvidou \cite{s4}.

We feel that to carry on the program of finding irreducible
representations of the history group for covariant systems
requires first a clarification of what constitutes the
history phase space of such a system, and how the
diffeomorphisms under which the system is covariant act on
this space and on the action functional.
Such a clarification was offered by one of us \cite{kk-diff} for
an arbitrary covariant finite-dimensional system. The
underlying diffeomorphisms are those on a one-dimensional
time manifold $\cal T$. A diffeomorphism-invariant
Lagrangian action of the system turns its configuration
space into a Finsler space. A parametrized Newtonian system
or a relativistic particle are just particular examples of
this general scheme. Let us summarize the main points which
emerged from this study:

\begin{itemize}
\item Finite-dimensional systems covariant under temporal
diffeomorphisms Diff$\cal T$ are described by means of two
{\it different} but {\it related} phase spaces:
\begin{enumerate}
\item an instantaneous phase space ${\cal G}_0\,$,
\item the phase space of histories $\cal G \,$.
\end{enumerate}
The instantaneous space ${\cal G}_0$ is the cotangent bundle
over the configuration space $\cal Q$. The phase space of
histories $\cal G$ includes not only the configuration space
histories and their conjugate momenta, but also another pair
of conjugate variables, the virtual metric of the time
manifold $\cal T$ and its conjugate momentum.
\item The symplectic form on ${\cal G}_0$ is induced by the
symplectic form on $\cal G\,$: the functions ${\cal F}({\cal
G}_0)$ on ${\cal G}_0$ (instantaneous dynamical variables)
can be mapped into ultralocal functionals ${\cal F}_{\rm UL}
({\cal G})\subset {\cal F} ({\cal G})$ on $\cal G$ in such a
way that the symplectic form on ${\cal G}_0$ is the pullback
of the symplectic form on $\cal G$ under this map.
\item The phase space of histories $\cal G$ carries a
representation of the group Diff$\cal T$ of temporal
diffeomorphisms by symplectomorphisms. Both the Lagrangian
and canonical history actions are invariant under Diff$\cal
T\,$.
\item Invariance of the history action implies that the
instantaneous canonical variables on ${\cal G}_0$ are
subject to a Hamiltonian constraint. This constraint,
however, does {\it not} generate time diffeomorphisms.
\end{itemize}

The present paper shows how to generalize these results to
covariant field theories. We explain the procedure for a
bosonic string moving in a Minkowskian target space $\cal
M\,$. This system is covariant under diffeomorphisms
Diff$\cal Y$ of a two-dimensional spacetime manifold $\cal
Y$ (the two-sheet). We chose this model for two reasons: It
is technically simpler than general relativity and, because
the quantum field theory of a bosonic string is
mathematically well defined, it seems feasible to build its
consistent histories projection operator formalism
version. General relativistic vacuum gravity will be treated
in the following paper.

The configuration history variables of the bosonic string
are the mapping $\sf \Phi$ from $\cal Y$ to $\cal M$ and the
pseudo-Riemannian metric $\sf G$ on $\cal Y$ which plays the
role of a Lagrange multiplier (section 2). The standard difficulty 
in constructing a covariant canonical formalism is the need of
an auxiliary structure, a foliation of $\cal Y$ by spacelike
hypersurfaces. When introduced as a fixed element, this
structure breaks the invariance of canonical action under
spacetime diffeomorphisms. We resolve this difficulty by
handling the foliation as a freely variable entity in the
canonical history action. Formally, the foliation is
described by a {\it time map} $\sf T$ from $\cal Y$ to a
one-dimensional time manifold $\cal T\,$. We {\it posit} the canonical
history action on $\cal G$ rather than trying to {\it
derive} it by something resembling the Legendre dual
transformation from a Lagrangian action on the configuration
space. For us, the history action is the fundamental
starting point of the history operator formalism rather than
an entity which needs to be derived from something else. The
proof of the action is in producing the desired equations of
motion, and hence we merely check that the canonical history
action correctly captures the string dynamics. This is done
by showing that when we vary the momentum ${\sf \Pi}_{\sf
\Phi}(y)$ in the history action and use the resulting
equation to eliminate ${\sf \Pi}_{\sf \Phi}(y)\,$, the
canonical history action reduces to the standard
Lagrangian Polyakov action (section 3). Let us emphasize that the
history action does not contain any fixed external structure
on $\cal Y\,$: All the fields ${\sf \Gamma}(y)$ can be varied
and the resulting equations are valid field equations. In
particular, this is true of ${\sf T}(y)\,$.

The strategy of taking the canonical history action as a
starting point also avoids the primary constraints. In the
Dirac constraint quantization, such constraints prohibit the
state function to depend on the conjugate coordinates. We do
not want to prejudice the issue what role such coordinates
may play in the construction of the decoherence functional
and history projection operators. Thus, e.g., in the
Gell-Mann and Hartle path integral approach, the lapse
function (a quantity which is conjugate to a primary
constraint) is explicitly used for partitioning
configuration space histories into diffeomorphism-invariant
classes.

The history phase space $\cal G$ is formed by genuine {\it spacetime} 
fields ${\sf \Gamma}(y)= \big ( {\sf \Phi}(y), {\sf \Pi}_{\sf
\Phi}(y), {\sf G}(y), {\sf \Pi}_{\sf G}(y), {\sf T}(y), {\sf
\Pi}_{\sf T}(y) \big )\,$, NOT by a one-parameter family of
conjugate spatial fields. The variation of the canonical history 
action is cast in terms of the history brackets between the conjugate 
fields on $\cal G$ (section 4). 

Having shown that the canonical action is equivalent to the Polyakov 
action, we also show that it is equivalent to the Dirac-ADM 
(Arnowitt, Deser and Misner)
canonical action for the string model (sections 5 to 8). This 
requires several
steps whose context we want to explain before plunging into
technical details.

First, we need to complement the time map $\sf T$ by the
space map $\sf X$ from $\cal Y$ to a {\it space manifold}
$\cal X \,$. While $\sf T$ describes a {\it
foliation} of spacetime by instants, $\sf X$ describes a
{\it reference frame}, i.e., a congruence of reference
worldlines. Taken together, the maps $\sf T$ and $\sf X$ can
be visualized as what the founders of general relativity
called the {\it reference fluid}. The particles of the fluid
identify space points and clocks carried by them identify
instants of time. This fixes the reference frame and time
foliation (section 5). 

The concept of reference fluid goes back to
Einstein \cite{ein} who coined for it the charming term
`mollusc', and to Hilbert who, in the second of his famous
communications on foundations of physics \cite{hil}
formalized the idea that the coordinate system should be
realized by a fluid carrying clocks which keep a causal
time. In general relativity, the reference fluid is
considered as a tenuous material system whose back reaction
on geometry can be neglected. There is just enough matter to
tell us where we are but not enough of it to disturb the
geometry. In this sense, the variables $\sf T$ and $\sf X$
are not dynamical.

Similar to $\sf T$, $\sf X$ is accompanied by a conjugate
history momentum ${\sf \Pi}_{\sf X}$, and the history phase
space $\cal G$ is extended by the pair ${\sf X},\, {\sf
\Pi}_{\sf X}\,$. Like the previous variables, the newly
introduced variables can also be freely varied. 
The transition from the canonical history action to the
Dirac-ADM action is carried on the extended space. The
formal aspects of this process demonstrate how to
incorporate the notion of reference fluid in the framework
of variational principles.\footnote{Though freely variable
in the action, the variables $\sf T$ and $\sf X$ which
describe the reference fluid do not affect the truly
dynamical variables (like $\sf \Phi$ in our model or the
gravitational field in general relativity). There is a
procedure which turns the reference fluid into a gravitating
physical system \cite{refflu1}-\cite{refflu4}; in the
simplest case, $\sf T$ and $\sf X$ become dynamical
variables describing a gravitating incoherent
dust. \cite{bk}}

The mappings $\sf T$ and $\sf X$ allow us to pull all the
remaining spacetime fields ${\sf \Phi}(y),\, {\sf \Pi}_{\sf
\Phi}(y)$ and $ {\sf G}(y), {\sf \Pi}_{\sf G}(y)$ back to the
${\cal T} \times \cal X$ manifold. They also induce the
familiar split of the pulledback metric into lapse, shift,
and the metric on the leaves of the time foliation. The
canonical history action can be expressed solely in terms of
the pulledback variables $\Phi (t,x),\,
{\Pi}_{\Phi}(t,x),\, G(t,x),\, {\Pi}_{G}(t,x)$ and their
temporal and spatial derivatives, and shown (with a caveat
to be discussed later) to coincide with the Dirac-ADM
action (section 6).

Symplectic structure of the history phase space allows us to
view this rearrangement of the action in a new light,
namely, as a canonical transformation to a new set of
conjugate variables on the extended history phase space. In
other words, our canonical history action and the Dirac-ADM
action are simply the same action on the extended history
phase space expressed merely in two different symplectic
charts (section 7).

The last point to understand is how the history Poisson
brackets induce the standard instantaneous brackets of the
Dirac-ADM dynamical variables ${\Phi} (t,x)$ and $
P_{\Phi}(t,x)\,$. As for finite-dimensional systems, the
{\it sine qua non} of this transition is the temporal
ultralocality of the Hamiltonian functional in the action.
An essential detail is that the Dirac-ADM momentum
$P_{\Phi}(t,x)$ is not simply the history momentum
$\Pi_{\Phi}(t,x)\,$, but the history momentum scaled down to
a temporal scalar by the lapse function (section 8). 

This is the caveat we have mentioned earlier: the symplectic chart 
provided by
the Dirac-ADM variables on the instantaneous phase spaces
${\cal G}_0$ becomes a non-symplectic chart on the history
phase space $\cal G\,$, and conversely, the symplectic chart
on the history phase space $\cal G$ induces a non-symplectic
chart on the instantaneous phase space ${\cal G}_0\,$. We
have discussed the geometric meaning of this feature in our
study of covariant finite-dimensional systems. \cite{kk-diff}

Technical details of the transition from the canonical
history action to the Dirac-ADM action span over sections
5 to 8 of this paper and occupy about one third of its
pages. We could never win our argument ad hominem in favor
of the canonical history action without convincing the
reader that, when properly understood, it is the same as the
familiar Dirac-ADM action. Let us emphasize that when it
comes to constructing a consistent histories projection
operator approach to the bosonic string, we advocate relying
entirely on the canonical history action and forgetting its
Dirac-ADM form.

We finally describe what we consider to be the central
result of our work. The canonical history formalism which we
propose is based on three manifolds, the spacetime $\cal Y$,
time $\cal T$, and (when it comes to the Dirac-ADM
formalism) the space $\cal X\,$. Each of these manifolds is
subject to its own group of diffeomorphisms. Our field
variables include the mappings between these manifolds and
between them and the target space $\cal M\,$. We claim that
our formalism is invariant under all of these
diffeomorphisms and that each of them is implemented by a
symplectomorphism in the history phase space $\cal
G\,$. This is the content of sections 9 to 11.

First, we show how the diffeomorphisms act on all the
configuration maps and their conjugate momenta (section 9). 
Second, we prove that all forms of the action $S$, the Lagrangian
Polyakov action, the canonical history action, and the
Dirac-ADM action, are invariant under all diffeomorphisms in
Diff$\cal Y$, Diff$\cal T$, and Diff$\cal X$ (section 10). 
Third, we
show that all these diffeomorphisms act as
symplectomorphisms on the history phase space $\cal G$ and
explicitly represent their generators by functionals in
${\cal F}({\cal G})$ subject to the history Poisson algebra (section 11).
\footnote{In \cite{ik}, spacetime diffeomorphisms were
implemented in canonical gravity by using the ordinary
instantaneous phase space. We now believe that the history
phase space $\cal G$ is the natural carrier of
representations of Diff$\cal Y\,$.}

Let us emphasize several features of our
construction. First, the diffeomorphism group acts on {\it
all} histories, i.e., on virtual histories as well as on
actual histories (which satisfy the equations of
motion). Second, the action functional is invariant on all
histories and even for such diffeomorphisms which move the
boundaries of the domain $\cal D$ in which the action is
varied to yield the equations of motion. The history phase
space on which the diffeomorphisms act is completely
different from the phase space of solutions of the equations
of motion. Third, the functionals on $\cal G$ which
represent Diff$\cal Y$ have nothing to do with the
Hamiltonian and momentum constraints on the data in ${\cal
G}_0\,$. Our examination of the status of spacetime
diffeomorphisms in canonical description of covariant
systems gives no support to a commonly held belief that
spacetime diffeomorphisms are generated by the Hamiltonian
and momentum constraints.
%%%%%%%%%%%%%%%%%%%%%%%%%%%%%%%%%%%%%%%%%%%%%%%%%%%%%%%%%%%

\section{Polyakov Action for a Bosonic String}

As our model of a generally covariant system we take a bosonic string. This is described by the Polyakov action
\begin{equation}
{S[\F,\G]} = -{1 \over 2} {\int} d^2y \;  \Big(  {|{\G}|}^{1 \over 2} \, {\iGab} \, {\eta}_{AB} \, {\FA}_{,\alpha} \, {\FB}_{,\beta}  \Big)(y)
\label{eq:ST-L}
\end{equation}
on a two-dimensional {\it spacetime manifold} $\yman$ whose elements $y \in \yman$ are called {\it events}. The coordinates of the event $y$ on $\yman$ are $y^{\alpha}$ where $\alpha = 0 , 1$.

The action depends on the metric $\G$ on $\yman$ represented by the components $\Gab$. The absolute value of the determinant of $\Gab$ is denoted by $|\G|$. The action also depends on the mapping
\begin{equation}
{\F} \, : \, {\yman} \, \rightarrow \, {\mman} \, \, \, \, {\rm by} \, \, \, \, y \in {\yman} \, \mapsto \, \phi = {\F}(y) \in {\mman}
\label{eq:introd of F}
\end{equation}
of $\yman$ to the {\it target space} $\mman$ which is an n-dimensional Minkowskian spacetime carrying the fixed flat metric ${\eta}$. In Minkowskian coordinates ${\phi}^A$, $A=0,1,...,n-1$, the mapping $\F$ is represented by a set of scalar fields $\FA$ and the metric $\eta$ by ${\eta}_{AB} = {\rm diag}(-1,1,...,1)$.

We shall limit our attention to closed strings for which $\yman$ has the topology of ${\rm I} \! {\rm R} \times {\rm S}^1$ and fix the temporal orientation of $\yman$. A variation $\delta{\F}$ of the map $\F$ is represented by the field ${\delta}{\FA}$ on $\yman$ consisting of vectors ${\delta}{\FA}$ on $\mman$ at points ${\phi}={\F}(y)$. Similarly, the variation $\delta{\G}$ of $\G$ is represented by the field ${\delta}{\Gab}$ of symmetric tensors on $\yman$. We say that $\delta{\F}$ is a variation of $\F$ in a compact domain ${\cal D} \subset \yman$ if ${\delta}{\FA}$ vanishes outside  ${\cal D}$. We say that  $\F$ has {\it fixed ends} on the boundary ${\cal B}={\partial}{\cal D}$ of  ${\cal D}$ if ${\delta}{\FA}$ is continuous on $\yman$. Otherwise, we say that $\F$ has {\it free ends}. For a closed string, ${\cal D}$ may be a compact spacetime region bounded by two spacelike hypersurfaces ${\cal B}_{(1)}$ and ${\cal B}_{(2)}$, ${\cal B} := {\partial}{\cal D} = {\cal B}_{(1)} \cup {\cal B}_{(2)}$, with ${\cal B}_{(2)}$ in the future of ${\cal B}_{(1)}$.

A field ${\F}$, ${\G}$ on $\yman$ is called a {\it virtual history}. An {\it actual history} is a field at which the action is stationary in comparison with all neighboring virtual histories obtained by varying $\G$ with free ends and $\F$ with fixed ends on any compact domain $\cal D$.
The variation of the action with respect to the mapping $\F$ which is fixed at the boundary ${\cal B}$ yields the Euler-Lagrange equations
\begin{equation}
{ {\delta S}  \over {\delta {\FA}(y)}} =  \Big( {\eta}_{AB} \, {|{\G}|}^{1 \over 2} \, {\iGab} \, {\bigtriangledown}_{\alpha} {\bigtriangledown}_{\beta} {\FB} \Big)(y)  = 0 \; .
\label{eq:ST-equs}
\end{equation}
The nabla operator ${\bigtriangledown}$ denotes the covariant derivative compatible with the metric $\G$,
\begin{equation}
{\bigtriangledown}_{\mu} {\Gab}(y) = 0  \; .
\label{eq:compatible}
\end{equation}
The spacetime metric $\G$ enters the action (\ref{eq:ST-L}) as a Lagrange multiplier and therefore it does not need to be fixed at ${\cal B}$. 
The variation with respect to $\G$ yields the energy-momentum tensor constraint
\begin{eqnarray}
\!\!\!\!\!\!\!\!\!\!\!\!\!\!\!\!\!\!\!\!\!\!\!\!\!\!\! {\sf T}^{\alpha\beta}(y) \!\!\!\! &:=& \!\!\!\! -2  {|{\G}(y)|}^{-{1 \over 2}} \, { {\delta S}  \over { \delta {\Gab}(y)} }
\nonumber
\\
&=& \!\!\!\! \Big( {\eta}_{AB} {\FA}_{,\mu} {\FB}_{,\nu} {\iGam} {\iGbn} - {1 \over 2} {\eta}_{AB} {\FA}_{,\mu} {\FB}_{,\nu} {\iGab} {\iGmn} \Big)(y) = 0 \; .
\label{eq:ST-cons}
\end{eqnarray}

\section{Canonical History Action on $\yman$}

Let us view the same system from a canonical spacetime perspective. This cannot be done without introducing on $\yman$ an auxiliary variable, the {\it time variable} $\sf T$. The time variable $\T$ is a mapping
\begin{equation}
{\T} \, : \, {\yman} \, \rightarrow \, {\tman} \, \, \, \, {\rm by} \, \, \, \, y \in {\yman} \, \mapsto \, t = {\T}(y) \in {\tman}
\label{eq:intro of T}
\end{equation}
from $\yman$ to a one-dimensional {\it time manifold} $\tman$ which has the topology of an open line ${\rm I} \! {\rm R}$. The elements $t \in \tman$ are called {\it moments}. The coordinate representation of the map $\T$ is $\T^0$ where $t^0$ is the coordinate of the moment $t \in {\tman}$. We require the gradient ${\Tt}_{,\alpha}$ of $\T$ to be timelike, $\iGab {\Tt}_{,\alpha} {\Tt}_{,\beta} < 0$. 
Each map $\T$ associates a spacelike hypersurface ${\Sigma}_{(t)}$ in $\yman$ with a point $t$ of $\tman$:
\begin{equation}
{\Sigma}_{(t)} = \Big\{ y {\in} {\yman} : {\T}(y) = t {\in} {\tman} \Big\} \; .
\label{eq:sigma}
\end{equation}
We call such a hypersurface {\it an instant} and their collection 
\begin{equation}
{\Sigma} = \{ {\Sigma}_{(t)} : t \in \tman \} 
\label{eq:time foliation}
\end{equation}
a {\it time foliation} ${\Sigma}$. We fix the orientation of $\tman$ and require that the time map $\T$ respect the orientation of $\yman$: if $t_1 \prec t_2$, the instant ${\Sigma}_{(t_2)}$ lies in the future of ${\Sigma}_{(t_1)}$ in $\yman$.

Still on the manifold $\yman$, we prescribe the {\it canonical history 
action} 
\begin{eqnarray}
{S[\F,\PF,\G,\T]} = {\int} d^2y \, \Big( {\PA} \, {\lie}{\FA} \Big)(y)
\nonumber
\\
- {1 \over 2} {\int} d^2y \, \Big(
{|\G|}^{-{1 \over 2}} \, {\eta}^{AB} \, {\PA} \, {\PB} + {|\G|}^{1 \over 2} \, {\Qab} \, {\eta}_{AB} \, {\FA}_{,\alpha} \, {\FB}_{,\beta} \Big)(y) \; .
\label{eq:ST-H}
\end{eqnarray} 
It depends on the previously introduced fields $\F$ and $\G$, on the time variable $\T$, and on the density $\PF$ (represented by the scalar densities $\PA$) considered as the momentum conjugate to $\F$. Let us emphasize that $\T$ is not a fixed structure but, like the remaining arguments of $S$, it can be freely varied. A particular choice of the variables $\F$, $\PF$, $\G$ and $\T$ is a {\it virtual history}.

Let us describe how the canonical action is constructed. The symbol $\lie$ denotes the Lie derivative with respect to a timelike unit vector field $\sf E$ normal to the foliation ${\Sigma}$ of $\yman$. This field is a vector concomitant\footnote{A tensor concomitant of a set of fields is a tensor constructed from the fields and their partial derivatives up to a finite order.} of the metric $\G$ and the time function $\T$:
\begin{equation}
{\Ea}[{\G},{\T}](y) := \left( - { {\iGab \, {\T}_{,\beta}} \over {\sqrt{-\iGmn \, {\T}_{,\mu} \, {\T}_{,\nu} }} } \right)(y)  \; .
\label{eq:e-field}
\end{equation}
Its normalization 
\begin{equation}
\Big( \sqrt{- \Gab \, \Ea \, \Eb} \, \Big)(y) = 1 
\label{eq:norm}
\end{equation}
follows directly from (\ref{eq:e-field}). The Lie derivative $\lie\F$ has the explicit form 
\begin{equation}
{\lie}{\FA}(y) = \Big( {\FA}_{,\alpha} \, {\Ea} \Big)(y) \; .
\label{eq:Lie}
\end{equation}
The tensor $\sf Q$,
\begin{equation}
{\Qab}[{\G},{\T}](y):= \Big( {\iGab} + {\Ea} \, {\Eb} \Big)(y) 
\label{eq:projector}
\end{equation}
is the projector orthogonal to the vector field $\sf E$, 
\begin{equation}
\Big( {\Ea} \, {\Gab} \, {\sf Q}^{\beta\gamma} \Big)(y) = 0 \; .
\label{eq:ortho}
\end{equation}
The first integral in (\ref{eq:ST-H}) is the Liouville term and the second one is the Hamiltonian term of the canonical action.

Before deriving the field equations, let us specify which variables need to be fixed on the boundary $\cal B$ of the region $\cal D$ and which can remain free. We know that the variable $\F$ must be fixed because its derivatives in the Lagrangian of (\ref{eq:ST-H}) would lead to an unwanted boundary term. On the other hand, the Lagrangian is an algebraic function of both $\G$ and $\PF$ and hence these variables can be left free on the boundary. Although the Lagrangian of (\ref{eq:ST-H}) depends on the derivatives of ${\T}$, we shall see that the ends of $\T$ can remain free.

By varying the spacetime action with respect to the field variables $\F$, $\PF$, $\G$ and $\T$, subject to the specified boundary conditions, we get the field equations which limit virtual histories to {actual histories}. The variation with respect to the dynamical fields $\F$ and $\PF$ yields the spacetime version of Hamilton's equations. The variation of $\PF$ generates the first set
\begin{equation}
{\lie\FA}(y) = \Big( \, {|{\G}|}^{-{1 \over 2}} \, {\eta}^{AB} \, {\PB} \Big)(y)  
\label{eq:H-equ1}
\end{equation}
which connects the momentum $\PF$ with changes of the field $\F$. The variation of $\F$ leads to the second set
\begin{equation}
{\lie}{\PA}(y) = \bigg( {|{\G}|}^{{1 \over 2}} \, {\eta}_{AB} \, {\bigtriangledown}_{\alpha} \Big( \Qab \, {\bigtriangledown}_{\beta} {\FB} \Big) \bigg)(y)
\label{eq:H-equ2}
\end{equation}
which determines the change of the momentum $\PF$.
As appropriate for a density, the Lie derivative of $\PF$ with respect to $\sf E$ is 
\begin{equation}
{\lie}{\PA}(y) = \Big( {\PA} \, {\Ea} \Big)_{,\alpha}(y) \; .
\label{eq:Lie2}
\end{equation}

When restricted to the configuration fields $\F$, the two sets of Hamilton
equations, (\ref{eq:H-equ1}) and (\ref{eq:H-equ2}), are equivalent to
the Euler-Lagrange equations (\ref{eq:ST-equs}). To show this, the first set (\ref{eq:H-equ1}) has to be solved for the momentum $\PF$,
\begin{equation}
{\PA}{[\F,\G,{\T}]}(y) = \Big( {|{\G}|}^{{1 \over 2}} \, {\eta}_{AB} \, {\lie}{\FB} \Big)(y) \; ,
\label{eq:solH-equ1}
\end{equation}
and the solution (\ref{eq:solH-equ1}) substituted in (\ref{eq:H-equ2}).
The Euler-Lagrange equations (\ref{eq:ST-equs}) then follow
by the use of (\ref{eq:projector}).

Instead of eliminating the momentum from the equations of motion, one can eliminate it from the action. In general, when one varies the action with respect to a subset of variables, solves the resulting equations for the variables which have been varied, and substitutes these solutions back in the action, one gets the reduced action that correctly determines the remaining variables. By applying this procedure to the variables $\PF$ and substituting the solutions (\ref{eq:solH-equ1}) back in the canonical action (\ref{eq:ST-H}), one reduces it to the Polyakov action (\ref{eq:ST-L}). The canonical history action (\ref{eq:ST-H}) is thus equivalent to the Polyakov action (\ref{eq:ST-L}) and we shall base on it the history approach to the bosonic string.

One cannot write the canonical action without introducing the time variable $\T$ but, when we eliminate the momentum $\PF$ from the canonical action, the time variable drops out of the Lagrangian of the reduced action (\ref{eq:ST-L}) as well. 
This allows us to anticipate that the field equation obtained by varying the canonical action with respect to $\T$ is redundant, being a consequence of the field equations (\ref{eq:H-equ1}) obtained by varying (\ref{eq:ST-H}) with respect to $\PF$. 
Let us verify this directly:

The variation of (\ref{eq:ST-H}) with respect to $\T$ yields 
\begin{equation}
{\delta}S=\int d^2y \, \bigg( \Big( {\PA} - {|\G|}^{1 \over 2} {\eta}_{AB} {\FB}_{,\beta} {\Eb}  \Big) \, {\FA}_{,\alpha} \, {\delta}{\Ea} \bigg)(y) \; ,
\label{eq:de}
\end{equation} 
where ${\delta}{\sf E}$ is induced by the variation ${\delta}{\T}$:
\begin{equation}
{\delta}{\Ea}(y) = \left( -{1 \over {\sqrt{-\iGmn \, {\Tt}_{,\mu} \, {\Tt}_{,\nu}}}}  \, {\Qab} \, {\partial}_{\beta}{\delta}{\Tt} \right)(y) \; .
\label{eq:dedt}
\end{equation} 
The variation (\ref{eq:de}) automatically vanishes due to the first set of Hamilton equations (\ref{eq:solH-equ1}). This means that the variation of $\T$ imposes no further conditions on the history fields and does not need to vanish on $\cal B$.

The variation of the action (\ref{eq:ST-H}) with respect to the metric $\G$ breaks into two contributions. There is a contribution of the form (\ref{eq:de})  where the variation ${\delta}{\sf E}$ is now induced by the variation ${\delta}{\G}$, 
\begin{equation}
{\delta}{\Ea}(y) = \bigg( \,  {1 \over 2} \,  \Big( {\Ea \, \Em \, \En - \Qam \, \En - \Qan \, \Em} \Big) \, {\delta}{\Gmn} \, \bigg)(y) \; .
\label{eq:dedg}
\end{equation} 
This contribution again vanishes because of (\ref{eq:solH-equ1}). The second contribution is obtained by varying (\ref{eq:ST-H}) with respect to $\G$ while keeping $\sf E$ fixed. It leads to the equation
\begin{eqnarray}
\!\!\!\!\!\!\!\!\!\!\!\!\!\!\!\!\!\!\!\! {\T}^{\mu\nu}(y) \!\!\!\! &:=& \!\!\!\! -2 {|{\G}(y)|}^{-{1 \over 2}} { {\partial S}  \over { \delta {\Gmn}(y) } } = \Big( \, {\eta}_{AB} {\FA}_{,\alpha} {\FB}_{,\beta} {\iGam} {\iGbn} \, \Big)(y)
\nonumber
\\
&+& \!\!\!\! \bigg( \, \, {1 \over 2} \, \Big( \, {|{\G}|}^{-1} \, {\eta}^{AB} \, {\PA} \, {\PB} - {\Qab} \, {\eta}_{AB} \, {\FA}_{,\alpha} \, {\FB}_{,\beta} \Big) \, {\iGmn} \, \bigg)(y) = 0
\label{eq:mixed cons 2}
\end{eqnarray} 
which, by virtue of (\ref{eq:solH-equ1}), is equivalent to the energy-momentum constraint (\ref{eq:ST-cons}).

\section{History Phase Space}

The variations of the canonical action (\ref{eq:ST-H}) can be accomplished by introducing the history Poisson brackets\footnote{History brackets were introduced by Isham and Linden within the history projection operator formalism \cite{il} and, since then, they have been widely used in the same context by them \cite{il2}, by Isham {\it et al} \cite{ilss}, by Savvidou \cite{s}, \cite{s2}, \cite{s4}, \cite{s5}, and by Anastopoulos and Savvidou \cite{sa}. They have been used by Kouletsis \cite{y0}, \cite{y} for reframing and generalizing the derivation of geometrodynamics from first principles (Hojman, Kucha\v{r} and Teitelboim \cite{hkt}) in the language of the history phase space.}
\begin{eqnarray}
\Big\{ \!\!\! \Big\{ \, {\FA}(y) \, , \, {\PB}(y') \, \Big\} \!\!\! \Big\} &=& {\delta}^A_B \, {\delta}(y,y') \; , 
\label{eq:ST-bra1} 
\\
\Big\{ \!\!\! \Big\{ \, {\Gab}(y) \, , \, {\Pgd}(y') \, \Big\} \!\!\! \Big\} &=& {\delta}^{\gamma\delta}_{\alpha\beta} \, {\delta}(y,y') \; , 
\label{eq:ST-bra2} 
\\
\Big\{ \!\!\! \Big\{ \, {\Tt}(y) \, , \, {\Pt}(y') \, \Big\} \!\!\! \Big\} &=& {\delta}^0_0 \, {\delta}(y,y') 
\label{eq:ST-bra3}
\end{eqnarray}
between conjugate canonical variables on the history phase space ${\cal G}$. The canonical fields 
\begin{equation}
{\sf \Gamma}(y) = \bigg( \, {\F}(y) \, , \, {\PF}(y) \, \, ; \, \, {\G}(y) \, , \, {\PG}(y) \, \, ; \, \, {\T}(y) \, , \, {\PT}(y)  \bigg) \in {\cal G}
\label{eq:phase space}
\end{equation}
whose components enter in (\ref{eq:ST-bra1})-(\ref{eq:ST-bra3}) are genuine spacetime fields. All the momenta are densities of weight 1 on $\yman$, because the delta function must transform as a scalar in the first argument but as a density of weight 1 in the second argument. The ${\delta}^A_B$ symbol is the Kronecker delta on the target space, ${\delta}^0_0=1$ is the Kronecker delta on $\tman$, and ${\delta}^{\gamma\delta}_{\alpha\beta}$ is the symmetrized product of two Kronecker deltas on $\yman$. Although the momenta $\PG$ and $\PT$ (represented by $\Pgd$ and $\Pt$) are needed to complement the coordinates $\G$ and $\T$ to the phase space ${\cal G}$, the canonical history action (\ref{eq:ST-H}) does not depend on these momenta. In particular, there are no corresponding Liouville-type terms in (\ref{eq:ST-H}). This is because $\G$ and $\T$ are not dynamical variables.

It is more appropriate to write the Poisson brackets (\ref{eq:ST-bra1})-(\ref{eq:ST-bra3}) in a smeared form. By smearing $\F$ and $\PF$ with the variations of the conjugate variables considered as external fields,
\begin{eqnarray}
{\F}[\delta{\PF}] := \int d^2y \, \Big( {\FA} \, {\delta}{\PA} \Big)(y) \; , 
\nonumber
\\
{\PF}[\delta{\F}] := \int d^2y \, \Big( {\PA} \, {\delta}{\FA} \Big)(y) \; , 
\label{eq:smeared}
\end{eqnarray}
the bracket relation (\ref{eq:ST-bra1}) can be replaced by 
\begin{equation}
\Big\{ \!\!\! \Big\{ \, {\F}[\delta{\PF}]  \, , \,  {\PF}[\delta{\F}]   \, \Big\} \!\!\! \Big\} = \int d^2y \, \Big( {\delta}{\PA} \, {\delta}{\FA} \Big)(y)   \; .
\label{eq:smeared equ1}
\end{equation}
Similar bracket relations can be written for the remaining canonical pairs when smearing them by the corresponding variations considered as external fields.

The history equations of motion amount to the condition that the Poisson brackets of the history action with all fundamental history variables (\ref{eq:phase space}) vanish: \footnote{This way of writing the variation of the action is due to Savvidou \cite{s} and Kouletsis \cite{y0}.}
\begin{equation}
\Big\{ \!\!\! \Big\{ \, {\sf \Gamma}(y) \, , \, {S[\F,\PF,\G,\T]} \, \Big\} \!\!\! \Big\} = 0 \; .
\label{eq:variation}
\end{equation} 
When we put ${\sf \Gamma}={\F}$, (\ref{eq:variation}) becomes
\begin{equation}
\int d^2y' \, \Big( {\lie\FC}  -  {|{\G}|}^{-{1 \over 2}}  \, {\eta}^{CB} \, {\PB}  \Big)(y')  \, \Big\{ \!\!\! \Big\{ \, {\FA}(y) \, , \, {\PC}(y') \, \Big\} \!\!\! \Big\}  = 0 \; .
\label{eq:p variation 1}
\end{equation}
By virtue of (\ref{eq:ST-bra1}), this amounts to the first set (\ref{eq:H-equ1}) of Hamilton equations.
Similarly, the choice ${\sf \Gamma}={\PF}$ leads to 
the second set (\ref{eq:H-equ2}) of Hamilton equations. 
More accurately, we should write these variations in the smeared form by requiring that
\begin{equation}
\Big\{ \!\!\! \Big\{ \,   {\PF}[\delta{\F}]    \, , \, {S[\F,\PF,\G,\T]} \, \Big\} \!\!\! \Big\} = 0 \; \; \; \; , \; \; \; \;
\Big\{ \!\!\! \Big\{ \,   {\F}[\delta{\PF}]    \, , \, {S[\F,\PF,\G,\T]} \, \Big\} \!\!\! \Big\} = 0 
\label{eq:variation of FP}
\end{equation} 
for all compact domains $\cal D$ and $\delta{\F}$ and $\delta{\PF}$ on these domains, where $\delta{\F}$ must be continuous on $\yman$. The continuity condition on the smearing function $\delta{\F}$ amounts to fixing the ends of $\F$. Notice that the domain $\cal D$ is associated with the smeared variables (\ref{eq:smeared}), not with the action functional $S$. 
Returning back to the unsmeared variation (\ref{eq:variation}), the choice ${\sf \Gamma}={\PG}$ leads to the constraint (\ref{eq:mixed cons 2}), while ${\sf \Gamma}={\PT}$ yields a superfluous equation satisfied as a consequence of 
(\ref{eq:H-equ1}). The brackets (\ref{eq:variation}) of the canonical action with the remaining variables ${\sf \Gamma}={\G}$ and ${\sf \Gamma} = {\T}$ vanish identically rather than imposing further conditions on the canonical fields.

The history field equations (\ref{eq:variation}) select from all virtual histories the actual ones that extremize the action. Notice that the set of equations (\ref{eq:variation}) is not preserved under general history brackets:
\begin{equation}
\Big\{ \!\!\! \Big\{ \, {\sf \Gamma}_{(2)}(y') \, , \, \Big\{ \!\!\! \Big\{ {\sf \Gamma}_{(1)}(y) \, , \, S{[\F,\PF,\G,\T]} \, \Big\} \!\!\! \Big\} \, \Big\} \!\!\! \Big\} \neq 0 \; .
\label{eq:2nd variation}
\end{equation}
Indeed, the left-hand side of (\ref{eq:2nd variation}) represents the second variations of the action which do not need to vanish.

\section{Pulling the Geometry Back to $\tman \times \xman$}

Our goal is to connect the canonical formalism on the phase space of histories ${\cal G}$ to the Dirac-ADM canonical formalism \cite{dirac}, \cite{adm} by reducing the spacetime canonical action (\ref{eq:ST-H}) to the Dirac-ADM canonical action. To do that, we need to have on $\yman$ not only a time variable $\T$ but also a space variable $\X$. The space variable $\X$ is a mapping 
\begin{equation}
{\sf X} \, : \, {\yman} \, \rightarrow \, {\xman} \, \, \, \, {\rm by} \, \, \, \, y \in {\yman} \, \mapsto \, x = {\sf X}(y) \in {\xman}
\label{eq:intro of X}
\end{equation}
from $\yman$ to a one-dimensional {\it space manifold} $\xman$ whose elements $x \in \xman$ are called {\it points}. The representation of the mapping $\X$ in terms of the coordinate $x^1$ on $\xman$ is denoted by $\Xx$. For a closed string, the space manifold $\xman$ has the topology of ${\rm S}^1$. We require the gradient ${\Xx}_{,\alpha}$ of $\sf X$ to be spacelike, $\iGab {\Xx}_{,\alpha} {\Xx}_{,\beta} > 0$. Each map $\sf X$ associates a timelike worldline $C_{(x)}$ in $\yman$ with a point $x$ of $\xman$:
\begin{equation}
C_{(x)} = \Big\{ y {\in} {\yman} : {\sf X}(y) = x {\in} {\xman} \Big\} \; .
\label{eq:timeline}
\end{equation}
We call such a worldline a {\it reference worldline} and their collection 
\begin{equation}
C = \{ C_{(x)} : x \in \xman \} 
\label{eq:reference frame}
\end{equation}
a {\it reference frame}.

To simplify further manipulations, let us introduce the product manifold $\zman = \tman \times \xman$, its points $z = (t,x) \in \zman$, and the local coordinates $z^a = (t^0, x^1)$ of $z$. The Cartesian product
\begin{eqnarray}
{\Z} := {\T} \times {\sf X} \, \, : \; \; \; \; {\yman} \, \, &\rightarrow& \, \, {\tman} \times {\xman} \, \, \, \, {\rm by} 
\nonumber
\\
y \in {\yman} \, \, &\mapsto&  \, \, \Big( t={\T}(y) \in {\tman} \, ,  \, x={\sf X}(y) \in {\xman}  \Big)
\label{eq:cartesian product}
\end{eqnarray}
of the time map $\T$ with the space map $\sf X$ tells us that the event $y \in \yman$ happened at the moment $t \in {\tman}$ and at the point $x \in {\xman}$. Its inverse 
\begin{eqnarray}
{Y} \, : \, {\tman} \times {\xman} \, \, \rightarrow \, \, {\yman} \, \, \, \, {\rm by}  \, \, \, \, \Big( t \in {\tman} \, , \, x \in {\xman} \Big) \, \, \mapsto  \, \, y=Y(t,x) \in {\yman}
\label{eq:foliation}
\end{eqnarray}
may be viewed as a one-parameter family $Y_{(t)}$, $t \in \tman$ of embeddings
\begin{equation}
{Y_{(t)}} \, : \, {\xman} \, \, \rightarrow \, \, {\yman} \, \, \, \, {\rm by}  \, \, \, \, x \in {\xman} \, \, \mapsto  \, \, {y_{(t)}} = {Y_{(t)}}(x) := Y(t,x) \in {\yman}
\label{eq:one-param}
\end{equation} 
of $\xman$ into $\yman$, whose images ${\Sigma}_{(t)}={Y_{(t)}}(\xman)$ define a foliation ${\Sigma}=\{ {\Sigma}_{(t)} , \, t \in \tman \}$ of $\yman$. It may also be viewed as a one-parameter family $Y_{(x)}$, $x \in \xman$ of curves
\begin{equation}
{Y_{(x)}} \, : \, {\tman} \, \, \rightarrow \, \, {\yman} \, \, \, \, {\rm by}  \, \, \, \, t \in {\tman} \, \, \mapsto  \, \, {y_{(x)}} = {Y_{(x)}}(t) := Y(t,x) \in {\yman} \; ,
\label{eq:comgruence}
\end{equation} 
whose images ${C}_{(x)}={Y_{(x)}}(\tman)$ define a reference frame $C=\{ {C}_{(x)} , x \in \xman \}$ in $\yman$. The mapping $Y$ locates the event $y \in \yman$ at which the instant ${\Sigma}_{(t)}=Y_{(t)}(\xman)$ intersects the reference worldline ${C}_{(x)}=Y_{(x)}(\tman)$. The inverse map to the embedding ${Y_{(t)}}$ is the restriction of the space map $\sf X$ to ${\Sigma}_{(t)}$:
\begin{equation}
{Y_{(t)}}^{-1} = {\sf X}{\mid}_{{\Sigma}_{(t)}} \, : \, {\Sigma}_{(t)} \, \, \rightarrow \, \, {\xman} \; .
\end{equation}
Similarly, the inverse map to the curve ${Y_{(x)}}$ is the restriction of the time map $\T$ to ${C}_{(x)}$: 
\begin{equation}
{Y_{(x)}}^{-1} = {\T}{\mid}_{{C}_{(x)}} \, : \, {C}_{(x)} \, \, \rightarrow \, \, {\tman}.
\end{equation}
Because the mappings $\Z$ and $Y$ are inverse to each other, we have 
\begin{equation}
{\Z} \circ Y = {\rm Id}_{\yman} \; \; \; \; {\rm and} \; \; \; \; Y \circ {\Z} = {\rm Id}_{\zman} \; .
\label{eq:Z map}
\end{equation}

The tensorial character of the differentiated mappings ${Y^{\alpha}}_{,a}$ and ${{\Z}^a}_{,\alpha}$ is captured by the positioning of the corresponding indices. The object ${Y^{\alpha}}_{,a}({\Z})$ can be considered as a basis of vectors on $\yman$ and ${{\Z}^a}_{,\alpha}$ as the dual cobasis. By differentiating (\ref{eq:Z map}), we get the orthonormality 
\begin{equation}
\Big( {{\Z}^a}_{,\alpha}(Y) \, {Y^{\alpha}}_{,b} \Big)(z)= {\delta}^a_b 
\label{eq:Orthonormality}
\end{equation}
and completeness 
\begin{equation}
\Big( {Y^{\alpha}}_{,a}({\Z})  \, {{\Z}^a}_{,\beta} \Big)(y)= {\delta}^{\alpha}_{\beta} 
\label{eq:Completeness}
\end{equation}
relations of the basis vectors. The Jacobian
\begin{equation}
\left| Y(z) \right| = {\rm det} \,  \Big( {Y^{\alpha}}_{,a}(z) \Big) = \Big( \, {1 \over 2} \, {\epsilon}_{\alpha\beta} \, {Y^{\alpha}}_{,a} \, {Y^{\beta}}_{,b} \, \, {\varepsilon}^{ab} \, \Big)(z)
\label{eq:JacobianY}
\end{equation}
of the mapping $Y$, written here in terms of the alternating tensor densities\footnote{The alternating symbols ${\sf \epsilon}_{\alpha\beta}$, ${\sf \epsilon}^{\alpha\beta}$ on $\yman$ and ${\varepsilon}_{ab}$, ${\varepsilon}^{ab}$ on $\zman$ are defined by their antisymmetry and by the property that ${\sf \epsilon}_{01}=1$, ${\sf \epsilon}^{01}=1$, ${\varepsilon}_{01}=1$, ${\varepsilon}^{01}=1$ in an arbitrary coordinate system on $\yman$ and $\zman$. The contravariant alternating symbols are densities of weight one and the covariant ones are densities of weight minus one on their respective spaces.} ${\sf \epsilon}_{\alpha\beta}$ on $\yman$ and ${\varepsilon}^{ab}$ on $\zman$, is a density of weight one on $\zman$ and of weight minus one on $\yman$. The Jacobian 
\begin{equation}
\left| {\Z}(y) \right| = {\rm det} \,  \Big( {{\Z}^a}_{,\alpha}(y) \Big) = \Big( \,  {1 \over 2} \, {\varepsilon}_{ab} \, {{\Z}^a}_{,\alpha} \, {{\Z}^b}_{,\beta} \, \,   {\epsilon}^{\alpha\beta} \, \Big)(y)
\label{eq:JacobianZ}
\end{equation}
of the mapping $\Z$ is a density of weight one on $\yman$ and of weight minus one on $\zman$. It holds that 
\begin{equation}
\left| Y(z) \right|  = \big| {\Z} \big( {Y}(z) \big) \big|^{-1} \; .
\label{eq:Jacobians}
\end{equation}

The mapping $Y$ can be considered as a functional $Y[{\Z}]$ of the mapping $\Z$. We can exhibit this dependence explicitly in the form of an integral 
\begin{equation}
Y^{\alpha}(z)[{\Z}] = \int d^2y \, \, y^{\alpha} \left| {\Z}(y) \right| \, {\delta}(z , {\Z}(y)) \; .
\label{eq:Y fnal Z}
\end{equation}
More generally, if ${\sf F}$ is a function of $y$, we can exhibit the functional dependence of $F[{\Z}] := {\sf F}(Y)$ on ${\Z}$ as
\begin{equation}
F(z)[{\Z}] = \int d^2y \, \, {\sf F}(y) \left| {\Z}(y) \right| \, {\delta}(z , {\Z}(y)) \; .
\label{eq:F fnal Z}
\end{equation}
The Jacobi matrix ${Y^{\alpha}}_{,a}$ can also be considered as a functional of ${\Z}$, ${Y^{\alpha}}_{,a}[{\Z}]$. By varying the identity (\ref{eq:Z map}),
\begin{equation}
Y^{\alpha} \big( {\Z}(y') \big) \big[ {\Z}\big] = y'^{\alpha} \; ,
\label{eq:varZmap}
\end{equation}
with respect to ${\Z}$, we learn that 
\begin{equation}
{    { {\delta}Y^{\alpha}(z)[{\Z}] } \over {{\delta}{\Z}^{a}(y) } } = - {Y^{\alpha}}_{,a}(z) \, \, {\delta}(Y(z),y) \; .
\label{eq:var Z map}
\end{equation}

The vector ${Y^{\alpha}}_{,1}$ is tangent to the instants ${\Sigma}_{(t)}$ while the {\it deformation vector} ${Y^{\alpha}}_{,0}$ is tangent to the worldlines ${C}_{(x)}$. The deformation vector is transverse but not necessarily orthogonal to ${\Sigma}_{(t)}$. It can be decomposed into components orthogonal and parallel to ${\Sigma}_{(t)}$:
\begin{equation}
{Y^{\alpha}}_{,0}(z) = \Big( {\N} \, e^{\alpha} + {\M} \, {Y^{\alpha}}_{,1} \Big)(z)   \; ,
\label{eq:lapse shift decomposition}
\end{equation}
where the unit normal 
\begin{equation}
e^{\alpha}(z) := {\Ea}(Y(z)) = \left( - { {\iGab {{\T}^0}_{,\beta}} \over {\sqrt{-\iGmn {{\T}^0}_{,\mu} {{\T}^0}_{,\nu} }} } \right)(Y(z))
\label{eq:small e-field}
\end{equation}
is both a temporal and a spatial scalar\footnote{In a one-dimensional manifold, like $\tman$ or $\xman$, tensors of $m$ upper indices and $n$ lower indices can be identified with scalar densities of weight $-m+n$. However, such an identification is not enforced here on the space manifold $\xman$ in order that the notation be generalizable to higher dimensions. The tensorial properties of fields on $\xman$ are captured by the positioning of the index $1$, but
their density character will not be exhibited in our notation and needs to be remembered.}.
The coefficient $N$ is called the {\it lapse function} and the coefficient $M$ the {\it shift vector}. This terminology reflects their tensorial character on $\xman$: The lapse is a spatial scalar and the shift a spatial vector. Both of them are temporal covectors, or temporal densities, on $\tman$. Their transformation properties on $\tman$ and $\xman$ are exhibited in our notation in (\ref{eq:lapse shift decomposition}) and (\ref{eq:small e-field}).

Let us now pull the spacetime metric $\G$ on $\yman$ back by the mapping $Y$ to the metric
\begin{equation}
{\gab}(z)[{\G},{\Z}] := \Big( {Y^{\alpha}}_{,a} \, {Y^{\beta}}_{,b} \, {\Gab}(Y) \Big)(z)
\label{eq:pull-backs}
\end{equation}
on $\zman$. 
The pullback ${\gxx}$ is the intrinsic (spatial) metric 
\begin{equation}
{\gxx}(z) =: {\qxx}(z) 
\label{eq:g11}
\end{equation}
of an instant ${\Sigma}_{(t)}$, i.e., $d{\sigma} = \sqrt{{\gxx}(t,x) \, dx^1 \, dx^1}$ is the intrinsic distance between two nearby events on ${\Sigma}_{(t)}$.\footnote{Notice that $Q_{11}$ is the only surviving component of the pullback of the projector ${\sf Q}_{\alpha\beta} = {\Gab} + {\sf E}_{\alpha} \, {\sf E}_{\beta}$ to $\zman$.} Similarly, the pullback ${\gtt}$ is the intrinsic (temporal) metric on a reference worldline ${C}_{(x)}$, i.e., $d{\tau} = \sqrt{-{\gtt}(t,x) \, dt^0 \, dt^0}$ is the proper time between two nearby events on ${C}_{(x)}$. The mixed component ${\gtx}$ of the pullback can be identified with the shift covector:
\begin{equation}
{\gtx}(z) = \Big( {\qxx} \, {\M} \Big)(z) \; . 
\label{eq:g01}
\end{equation}
The temporal metric ${\gtt}$ can be expressed as a combination of the lapse and shift coefficients,
\begin{equation}
{\gtt}(z) = \Big( - {\N} \, {\N} + {\qxx} \, {\M} \, {\M} \Big)(z) \; . 
\label{eq:g00}
\end{equation}

Similarly, let us push the contravariant metric ${\G}^{\alpha\beta}$ on $\yman$ forward to the contravariant metric
\begin{equation}
{\igab}(z)[{\G},{\Z}] :=  \Big( {{\Z}^a}_{,\alpha} \, {{\Z}^b}_{,\beta} \, {\iGab} \Big)(Y(z))
\label{eq:contra G}
\end{equation}
on $\zman$. By taking the inverse of the covariant metric ${\gab}$ in (\ref{eq:g11})-(\ref{eq:g00}), we learn that 
\begin{eqnarray} 
{\igtt}(z) = \left( - {1 \over {\N}} {1 \over {\N}} \right)(z) &,&  {\igtx}(z) = \left( - {1 \over {\N}} \, {{\M} \over {\N}} \right)(z)  \; , 
\nonumber
\\
{\igxx}(z) \!\!\! &=& \!\!\! \left(  Q^{11} - {{\M} \over {\N}} \, {{\M} \over {\N}} \right)(z) \; ,
\label{eq:ggg01}
\end{eqnarray}
where ${Q^{11}}=({\qxx})^{-1}$ is the inverse of the covariant spatial metric ${Q}_{11}$. These equations allow us to express the lapse $N$, the shift $M$ and the spatial metric $Q$ in terms of the components ${\G}^{\alpha\beta}$ of the contravariant metric,
\begin{eqnarray}
{\N}(z) =  (-{\igtt})^{- {1 \over 2}}(z) &,& {\M}(z) = \left( - {{\igtx} \over {\igtt}} \right)(z) \; , \; 
\nonumber
\\
{\qxx}(z) \!\!\! &=& \!\!\! \left( {\igxx} + { {{\igtx} \, {\igtx} \over {\igtt} }} \right)^{-1}(z) \; ,
\label{eq:final expressions}
\end{eqnarray}
and hence, through (\ref{eq:contra G}) and (\ref{eq:F fnal Z}), exhibit explicitly their functional dependence on $\G$ and $\Z$.

\section{Rewriting The Canonical History Action}

After pulling the spacetime metric $\G$ back to $\zman = \tman \times \xman$, let us do the same thing to the dynamical variables $\F$ and $\PF$:
\begin{eqnarray}
{\fa}(z)[{\F},{\Z}] & := &  {\FA} (Y(z)) \; ,  
\label{eq:pull-back}
\\
{\pa}(z)[{\PF},{\Z}] &:= & \left| Y(z) \right| \, {\PA}(Y(z)) \; .
\label{eq:density mom}
\end{eqnarray}
When we write the Jacobian $\left| Y \right|$ in the form (\ref{eq:JacobianZ})-(\ref{eq:Jacobians}), the substitution rule (\ref{eq:F fnal Z}) explicitly exhibits the dynamical variables (\ref{eq:pull-back})-(\ref{eq:density mom}) as functionals of $\Z$. The history momentum $\pf$ in (\ref{eq:density mom}) is both a temporal and a spatial density. Let us also scale $\pf$ down by the lapse function into a temporal scalar $P_{\f}$,
\begin{equation}
{P_A}(z)[{\PF},{\G},{\Z}] :=  {\N}^{-1}(z)[{\G},{\Z}] \, {\pa}(z)[{\PF},{\Z}] \; ,
\label{eq:ADM mom}
\end{equation}
which will play the role of momentum in the standard Dirac-ADM formalism. The variable $P_{\f}$ is still a spatial density.\footnote{Further on, we will also introduce the momentum $\pg$ conjugate to the metric $\g$ and identify the momenta $P_N$ and $P_M$ conjugate to the lapse $N$ and the shift $M$. Our notation will consistently stick to the rule that all quantities denoted by $\Pi$ or $P$ are spatial densities; those denoted by $\Pi$ are also temporal densities, while those denoted by $P$ are temporal scalars.}

Let us now pull the history action (\ref{eq:ST-H}) back to $\tman \times \xman$ by the mapping $Y$ and express it via the dynamical fields $\f$ and $P_{\f}$ and the metric coefficients $Q$, $N$ and $M$. However, we still consider the action as a functional of the original history variables ${\F}$, ${\PF}$, ${\G}$ and ${\T}$. We get
\begin{equation}
S[{\F},{\PF},{\G},{\T}] = \int dt^0 \int dx \; \Big( {P_A} \, {{\Phi}^A}_{,0} 
- {\M} H_{1} - {\N} H \Big)(z) \; ,
\label{eq:ADM action2}
\end{equation}
where $H_{1}$ denotes the supermomentum 
\begin{equation}
H_{1}(z) = \Big( {P_A} \, {{\Phi}^A}_{,1} \Big)(z)
\label{eq:supermom}
\end{equation}
and ${H}$ the super-Hamiltonian 
\begin{equation}
{H}(z) = \Big( \, {1 \over 2} \, |Q|^{-{1 \over 2}} \, {\eta}^{AB} \, P_A \, P_B +  {1 \over 2} \, |Q|^{1 \over 2} \, {\eta}_{AB} \, {Q^{11}} \, {{\Phi}^A}_{,1} \, {{\Phi}^B}_{,1} \, \Big)(z) \; .
\label{eq:superham}
\end{equation}
The square root $|Q|^{{1 \over 2}}$ of the determinant $|Q|$ is a spatial scalar density. By their construction, (\ref{eq:supermom})-(\ref{eq:superham}), the supermomentum $H_1$ and the super-Hamiltonian $H$ are temporal scalars but spatial densities, similarly as $P_{\f}$.  This makes the integral (\ref{eq:ADM action2}) invariantly defined on $\zman = \tman \times \xman$.

Let us spell out the steps that lead from (\ref{eq:ST-H}) to (\ref{eq:ADM action2}). The first two terms in (\ref{eq:ADM action2}) stem from the Lie derivative term in the action (\ref{eq:ST-H}). This becomes clear when we rewrite the lapse-shift decomposition (\ref{eq:lapse shift decomposition}) as an equation for $e$,
\begin{equation}
e^{\alpha}(z) := {\Ea}(Y(z)) = {\N}^{-1}(z) \, \Big( \, {Y^{\alpha}}_{,0} \, - \, {\M} \, {Y^{\alpha}}_{,{1}} \Big)(z) \; ,
\label{eq:e-decompose}
\end{equation}
express the Lie derivative as
\begin{equation}
({\lie\FA})(Y(z))= {\N}^{-1}(z) \, \Big( \, {{\Phi}^A}_{,0} \, - \, {\M} \, {{\Phi}^A}_{,1} \, \Big)(z) \; ,
\label{eq:Lie exp}
\end{equation} 
and use the definition (\ref{eq:ADM mom}) of $P_{\f}$. 
The super-Hamiltonian term in (\ref{eq:ADM action2}) arises directly from the Hamiltonian in (\ref{eq:ST-H}) through the relations (\ref{eq:final expressions}), (\ref{eq:pull-back})  and the definition (\ref{eq:ADM mom}) of $P_{\f}$. 

Let us stress that the action (\ref{eq:ADM action2}) is exactly the same functional of the spacetime variables ${\F}$, ${\PF}$, $\G$ and ${\T}$ as the action (\ref{eq:ST-H}), and that it does not depend on the space map ${\sf X}$. The same thing is true about the equations of motion. The space map ${\sf X}$ is needed only for introducing the intermediate variables $\f$, $\pf$ (or $P_{\f}$) and $\g$ (or $Q$, $N$ and $M$) and writing the action as an integral over $\zman = \tman \times \xman$.

The introduction of $\sf X$ is naturally followed by the corresponding extension of the history phase space ${\cal G}$. Its elements ${\sf \Gamma}$ are now the conjugate fields
\begin{equation}
{\sf \Gamma}(y) = \bigg( \, {\F}(y) \, , \, {\PF}(y) \, \, ; \, \, {\G}(y) \, , \, {\PG}(y) \, \, ; \, \, {\Z}(y) \, , \, {\PZ}(y)  \, \bigg) \in {\cal G} \; .
\label{eq:ext phase space}
\end{equation}
Like its counterpart $\T$, the space map ${\X}$ is complemented by the conjugate momentum ${\PX}$ with the history bracket
\begin{equation}
\Big\{ \!\!\! \Big\{ \,  {\sf X}^{1}(y) \, , \, {\Px}(y') \,  \Big\} \!\!\! \Big\} = {\delta}^1_1 \, {\delta}(y,y') 
\label{eq:ST-bra4}
\end{equation}
which is adjoined to the previous set of brackets (\ref{eq:ST-bra1})-(\ref{eq:ST-bra3}). The trivial additional variations of the action (\ref{eq:ST-H}) yield the identities
\begin{equation}
\Big\{ \!\!\! \Big\{ \, {\sf X}^{1} (y) \, , \, S \, \Big\} \!\!\!\Big\} \equiv 0 \; \; , \; \; \Big\{ \!\!\! \Big\{ \, {\Px}(y) \, , \, S \, \Big\} \!\!\! \Big\} \equiv 0 \; .
\label{eq:trivial 12}
\end{equation} 
The canonical coordinates ${\Z} = ({\T} , {\X})$ and the momenta ${\PZ} = ({\PT}, {\PX})$ collect the brackets (\ref{eq:ST-bra3}) and (\ref{eq:ST-bra4}) into a single equation
\begin{equation}
\Big\{ \!\!\! \Big\{ \, {\Za}(y) \, , \, {\Pb}(y') \, \Big\} \!\!\! \Big\} = {\delta}^a_b \, {\delta}(y,y') \; .
\label{eq:ST-braZ}
\end{equation}

\section{Performing the Canonical Transformation}

We now take a key step towards turning the canonical history action into the Dirac-ADM form: We show that there exists a canonical transformation on the history phase space (\ref{eq:ext phase space}) which turns the variables $\f$, $\pf$ and $Q$, $N$ and $M$ on $\zman$ into new fundamental canonical variables on ${\cal G}$. More specifically, our canonical transformation is performed in two steps. The first step takes the old canonical variables (\ref{eq:ext phase space}) which are tensor fields on $\yman$, and transforms them into new canonical variables
\begin{equation}
\Bigg( \, {\Gamma}(z) = \bigg( \, {\f}(z) \, , \, {\pf}(z) \, \, ; \, \, {\g}(z) \, , \, {\pg}(z) \, \bigg) \, \, ; \, \, {\bar{\Z}}(y) \, , \, {\bar {\sf \Pi}}_{\bar {\Z}}(y) \, \Bigg) \in {\cal G} 
\label{eq:new variables}
\end{equation}
such that the first two pairs become fields ${\Gamma}$ on $\zman = \tman \times \xman$, while the third pair still retains its character of fields on $\yman$.\footnote{This type of canonical transformation was introduced by Brown and Kucha\v{r} \cite{bk} in the context of the standard Dirac-ADM Poisson bracket. Here, it is generalized to the history framework.} The generating functional $F$ of this transformation is a functional of the new canonical coordinates ${\f}$, ${\g}$ and ${\bar{\Z}}$ and of the old canonical momenta ${\PF}$, ${\PG}$ and ${\PZ}$. It has the form
\begin{eqnarray}
&& F[ {\f} , {\g} , {\bar{\Z}} ; {\PF} , {\PG} , {\PZ} ] = 
\nonumber
\\
&& \int d^2y \, \Big( {\fa}(\bar{\Z}) \, {\PA}  + 
{\gab}(\bar{\Z})  \, \, {{{\bar{\Z}}^a}}{}_{,\alpha} \,  \, {{{\bar{\Z}}^b}}{}_{,\beta} \, {\Pab}  +  {\bar{\Z}}^a \,  \, {\sf \Pi}_a \Big)(y) \; ,
\label{eq:gen functional}
\end{eqnarray}
and it generates the canonical transformation according to the standard rules
\begin{eqnarray}
{\FA}(y) = { {\delta F} \over {\delta {\PA}(y)} }\, , &{}& \; \;   {\pa}(z) = { {\delta F} \over {\delta {\fa}(z)} } \; ,
\nonumber
\\
{\Gab}(y) = { {\delta F} \over {\delta {\Pab}(y)} }\, ,  &{}& \; \;   {\pab}(z) = { {\delta F} \over {\delta {\gab}(z)} } \; ,
\nonumber
\\
{\Z}^a(y) = { {\delta F} \over {\delta {\sf \Pi}_a(y)} }\, , &{}&  \; \;  {\bar {\sf \Pi}}_{a}(y) = { {\delta F} \over {\delta {\bar{\Z}}_a(y)} } \; .
\label{eq:standard rules}
\end{eqnarray}
Because $F$ is a linear functional of the momenta, it generates a point transformation on the history phase space $\cal G$, i.e., the new canonical coordinates depend only on the old coordinates, not on the momenta. However, because $F$ depends on the derivatives of the mapping $\Z$, the resulting point transformation is not ultralocal. When resolved with respect to the new canonical variables, the transformation equations (\ref{eq:standard rules}) read 
\begin{eqnarray}
{\fa}(z)[{\F} , {\Z}] &=& {\FA}(Y(z)) \; , 
\label{eq:transf1a}
\\
{\pa}(z)[{\PF} , {\Z}] &=& \left| Y(z) \right| \, {\PA}(Y(z)) \; , 
\label{eq:transf1b}
\\
{\gab}(z)[{\G} , {\Z}] &=& \Big( {Y^{\alpha}}_{,a} \, {Y^{\beta}}_{,b} \, {\Gab}(Y) \Big)(z) \; , 
\label{eq:transf2a}
\\
{\pab}(z)[{\PG} , {\Z}] &=& \left| Y(z) \right| \, \Big( {{\Z}^a}_{,\alpha} \, {{\Z}^b}_{,\beta} \, {\Pab} \Big)(Y(z)) \; , 
\label{eq:transf2b}
\\
\nonumber
\\
{\bar{\Z}}^a(y)[{\Z}] &=& {\Z}^a(y) \; , 
\label{eq:transf3}
\\
{\bar {\sf \Pi}}_{a}(y)[{\Z} , {\PZ} , {\F} , {\PF} , {\G} , {\PG}] &=&  \Big( {\sf \Pi}_a + {\PA} \, {\cal L}_{(Y_{,a})}{\FA} \, + {\Pab} \, {\cal L}_{(Y_{,a})}{\Gab} \Big)(y)  \; ,
\nonumber
\\
\label{eq:transf4}
\end{eqnarray}
where the Lie derivatives in (\ref{eq:transf4}) are taken along the two vector fields $a=0,1$ of the Jacobi matrix ${Y^{\alpha}}_{,a}(y)[{\Z}]$. The orthonormality conditions (\ref{eq:Orthonormality}) imply that ${\cal L}_{(Y_{,b})}{\Za} = {\delta}^a_b$, which explains the absence of a Lie derivative term multiplying $\Pa$ on the right-hand side of (\ref{eq:transf4}).

The first three equations in (\ref{eq:transf1a})-(\ref{eq:transf4}) reproduce the pullbacks (\ref{eq:pull-back}), (\ref{eq:density mom}), (\ref{eq:pull-backs}) of the fields $\F$, $\PF$  and $\G$ to $\zman$, while (\ref{eq:transf2b}) yields the pullback of the momentum $\PG$ to $\zman$. The canonical transformation does not affect the mapping ${\Z}$, (\ref{eq:transf3}). However, the canonical momentum ${\bar {\sf \Pi}}_{\bar {\Z}}$ conjugate to ${\bar{\Z}}={\Z}$ depends on all of the old canonical variables, (\ref{eq:transf4}). Its particular form will concern us again when we discuss the representation of infinitesimal diffeomorphisms of $\yman$, $\tman$ and $\xman$ by symplectomorphisms of ${\cal G}$.

The second, and final, step is to complete the decomposition of ${\g}$ into the lapse, the shift and the induced metric , 
\begin{eqnarray}
{\gxx}(z) = {\qxx}(z) &,&   {\gtx}(z) = \Big( {\qxx} \, {\M} \Big)(z) \; ,
\nonumber
\\
{\gtt}(z) = \Big(- {\N} \, {\N} \, \!\!\!  &+& \!\!\! \,  {\qxx} \, {\M} \, {\M} \,  \Big)(z) \; ,
\label{eq:equation1}
\end{eqnarray}
into another point transformation on the history space,
\begin{equation}
\Big( \; {\g}(z) \; , \; {\pg}(z) \; \Big) \; \;   \mapsto \;  \; \Big( \; Q(z) \; , \; {{\Pi}_Q}(z) \; \;  ; \;  \;  {N}(z) \;  , \;  {P_N}(z) \; \;   ; \;  \;   {M}(z) \;  , \;  {P_M}(z) \; \Big) \; .
\label{eq:ultra transf}
\end{equation}
Because (\ref{eq:equation1}) is ultralocal, this task reduces to solving (\ref{eq:equation1}) for the new canonical coordinates, which was done in (\ref{eq:final expressions}), and putting\footnote{The single component ${{\Pi}}^{11}$ of the momentum ${\Pi}_Q$ should not be confused with the $11$-component of the tensor $\pab$ representing the momentum ${\Pi}_G$. These two components are not equal, as we can see from the last equations in (\ref{eq:actual ultra transf}) and (\ref{eq:final ultra transf}). There is no occasion for being misled because we never need the individual components of $\pab$ in our equations.}
\begin{eqnarray}
{P}(z) = \left( { {\partial {\gab}} \over {\partial {\N}}  }  {\pab} \right)(z) \, \!\!\!\! &,& \!\!\!\! \, {P}_1(z) = \left( { {\partial {\gab}} \over {\partial {\M}}  }  {\pab} \right)(z) \, , 
\label{eq:actual ultra transf1}
\\
{{\Pi}}^{11}(z) \!\!\!\! &=& \!\!\!\! \left( { {\partial {\gab}} \over {\partial {Q_{11}}}  }  {\pab} \right)(z) \, .
\label{eq:actual ultra transf}
\end{eqnarray}
The canonical momentum ${{\Pi}_Q}$ is again a temporal density but the canonical momenta ${P_N}$ and ${P_M}$ are temporal scalars. Equations (\ref{eq:actual ultra transf1})-(\ref{eq:actual ultra transf}) can be brought to the final form
\begin{eqnarray}
\!\!\!\!\!\!\!\!\!\!\!\!\!\!\!\! {P}(z) = \Big( -2 (-G^{00})^{-{1 \over 2}} \, {\Pi}^{00} \Big)(z) &,& {P}_1(z) = \Big( 2 G_{1a} \, {\Pi}^{0a}  \Big)(z) \; , 
\label{eq:final ultra transf1}
\\
{{\Pi}}^{11}(z) \!\!\! &=& \!\!\! \Big( (G_{11})^{-2} \, G_{1a} \, G_{1b} \, {\pab} \Big)(z) \; ,
\label{eq:final ultra transf}
\end{eqnarray}
which completes our canonical transformation.

Because the transformation (\ref{eq:transf1a})-(\ref{eq:transf4}) from the old variables (\ref{eq:ext phase space}) to the new variables (\ref{eq:new variables}) is canonical, we can directly write the history Poisson brackets of the new variables. The only ones that do not vanish are, of course, those within the conjugate pairs $({\f},{\pf})$, $({\g},{\pg})$ and $({\bar {\Z}},{\bar {\sf \Pi}}_{\bar {\Z}})$,
\begin{eqnarray}
\Big\{ \!\!\! \Big\{ \, {\fa}(z) \, , \, {\pb}(z') \, \Big\} \!\!\! \Big\} &=& {\delta}^A_B \, {\delta}(z,z') \; , 
\label{eq:ind-bra1} 
\\
\Big\{ \!\!\! \Big\{ \, {\gab}(z) \, , \, {\Pi}^{cd}(z') \, \Big\} \!\!\! \Big\} &=& {\delta}^{cd}_{ab} \, {\delta}(z,z') \; ,
\label{eq:ind-bra2}  
\\
\Big\{ \!\!\! \Big\{ \, {\bar{\Z}}^a(y) \, , \, {\bar {\sf \Pi}}_{b}(y') \, \Big\} \!\!\! \Big\} &=& {\delta}^a_b \, {\delta}(y,y') \; .
\label{eq:ind-bra3}
\end{eqnarray}
Alternatively, after the second point transformation (\ref{eq:ultra transf}) of the metric variables, accomplished by (\ref{eq:final expressions}) and (\ref{eq:final ultra transf}), the geometric bracket (\ref{eq:ind-bra2}) is replaced by
\begin{eqnarray}
\Big\{ \!\!\! \Big\{ \, {\N}(t,x) \, , \, {P}(t',x') \, \Big\} \!\!\! \Big\} &=& {\delta}(t',t)_0 \,  {\delta}(x,x') 
\; , 
\label{eq:2ind-bra1} 
\\
\Big\{ \!\!\! \Big\{ \, {\M}(t,x) \, , \, {P}_1(t',x') \, \Big\} \!\!\! \Big\} &=& {\delta}^1_1 \, {\delta}(t',t)_0 \, {\delta}(x,x') \; ,
\label{eq:2ind-bra2}  
\\
\Big\{ \!\!\! \Big\{ \, Q_{11}(t,x) \, , \, {{\Pi}}^{11}(t',x') \, \Big\} \!\!\! \Big\} &=& {\delta}^{11}_{11} \, {\delta}(t,t') \, {\delta}(x,x') \; .
\label{eq:2ind-bra3}  
\end{eqnarray}
Notice that the order in which $t$ and $t'$ appear in the arguments of the temporal delta function in (\ref{eq:2ind-bra1})-(\ref{eq:2ind-bra2}) has been reversed, and that it no longer follows the order in which $t$ and $t'$ enter the Poisson brackets. In this case, we explicitly marked the temporal density character of the delta function in its second argument by the covector index $0$, so that it explicitly matches the temporal density character of the canonical coordinates $\N$ and $\M$ on the left sides of (\ref{eq:2ind-bra1})-(\ref{eq:2ind-bra2}).

\section{$\!\!\!$The History Action in the Dirac-ADM Form}

The history action (\ref{eq:ADM action2}) can now be considered as a functional of the {\it new} canonical history variables:
\begin{equation}
S[{\f} , {\pf} ; Q , {N} , {M}] = {\Theta} - {\cal H} \; .
\label{eq:ADM proper}
\end{equation}
We have split it into two distinctive pieces, the {\it Liouville functional}
\begin{equation}
{\Theta}[{\f} , {\pf} ; {N}] = \int dt^0 \int dx \; \Big( {P_A} \, {\fa}_{,0} \Big)(t,x) \; ,
\label{eq:Liouville fnal}
\end{equation}
and the {\it Hamiltonian functional}
\begin{equation}
{\cal H}[{\f},{\pf};Q,N,M] = \int dt^0 \int dx \; \Big( {\N} \, H + {\M} \, H_1 \Big)(t,x) \; ,
\label{eq:Hamiltonian Fnal}
\end{equation}
where the supermomentum $H_1$ and super-Hamiltonian $H$ in (\ref{eq:Hamiltonian Fnal}) are given in terms of ${\pf}$ by (\ref{eq:supermom}) and (\ref{eq:superham}) through the intermediary variable (\ref{eq:ADM mom}),
\begin{equation}
P_{\f} = {N}^{-1} {\pf}  \; .
\label{eq:intermediary}
\end{equation}
Notice that ${\Theta}$ is also expressed through the intermediary $P_{\f}$ in (\ref{eq:Liouville fnal}).

From the history point of view, the Liouville functional $\Theta$ is characteristically different from the Hamiltonian functional ${\cal H}$. The Liouville functional is {\it spatially ultralocal}, i.e., its integrand does not depend on the spatial derivatives of the canonical variables. On the other hand, the Hamiltonian functional is {\it temporally ultralocal}, i.e., its integrand does not depend on the temporal derivatives of the canonical variables. It is the choice of the new history variables which leads to this characteristic space-time split. Also notice that the history action written in the new variables does not depend at all on the time and space maps ${\bar{\Z}}^a$. This distinguishes it from the canonical history action (\ref{eq:ST-H}) or (\ref{eq:ADM action2}) which depends on the time map $\T$ (though it also does not depend on the space map $\sf X$). 

Because ${\Theta}$, $H_1$ and $H$ are given through the intermediary variable $P_{\f}$ they depend explicitly on the lapse function $N$. But when viewed as functionals of the variables $P_{\f} = N^{-1} {\pf}$ they all lose their dependence on $N$, and the action (\ref{eq:ADM proper}) becomes the familiar Dirac-ADM action. The calculus of variations is indifferent to whether it is the pair $P_{\f}$, $N$ or the pair $\pf$, $N$ that is independently varied. However, when we take the Dirac-ADM choice, the Hamiltonian functional ${\cal H}$ in (\ref{eq:Hamiltonian Fnal}) becomes a {\it linear} functional of $N$ and $M$, 
\begin{equation}
H[{\f},{P_{\f}};Q,N,M] = \int dt^0 \int dx \; \Big( {\N} \, H + {\M} \, H_1 \Big)(t,x) = H[N] + H[M] \; ,
\label{eq:linear Hamiltonian Fnal}
\end{equation}
where $H_1$ and $H$ in (\ref{eq:linear Hamiltonian Fnal}) are now given directly by (\ref{eq:supermom}) and (\ref{eq:superham}) in terms of the Dirac-ADM momentum ${P_{\f}}$. The linearity in the lapse and shift is underlined in the notation $H[N]$ and $H[M]$ of (\ref{eq:linear Hamiltonian Fnal}) which singles out these variables as {\it smearing functions}.

The variation of the Dirac-ADM action (\ref{eq:linear Hamiltonian Fnal}) with respect to the dynamical variables $\f$ and $P_{\f}$ leads to the ordinary Hamilton equations which can be expressed via the standard instantaneous brackets, and its variation with respect to the multipliers $N$, $M$ and $Q$ imposes constraints on the instantaneous data. The simple change of temporal weight (\ref{eq:intermediary}) thus accomplishes the transition from the history variables and Poisson brackets to their instantaneous counterparts. Without that change, the transition cannot be accomplished. Let us therefore show in detail how the history Poisson brackets on the phase space ${\cal G}$ {\it induce} the standard Dirac-ADM instantaneous Poisson brackets.

To begin with, let us emphasize that $P_{A}$ introduced by (\ref{eq:ADM mom}) is {\it not} a history momentum conjugate to $\fa$. Indeed,
\begin{equation}
\Big\{ \!\!\! \Big\{ \, {\fa}(t,x)     \, , \,  P_{B}(t',x')    \, \Big\} \!\!\! \Big\} = {\N}^{-1}(t',x') \, {\delta}(t,t')_0 \, {\delta}(x,x') \; .
\label{eq:non bracket}
\end{equation}
The canonical variable ${\N}^{-1}$ on the right-hand side of (\ref{eq:non bracket}) is needed to turn the temporal delta function into a temporal biscalar, and its presence there tells us that the chart of coordinates including $P_{A}$ is not a canonical chart.\footnote{This distinguishes $P_{\f}$ from the variables $P_N$ and $P_M$ which are {\it bona fide} history momenta conjugate to the canonical coordinates $N$ and $M$.} However, we can obtain the Dirac-ADM equations from the {history action} (\ref{eq:ADM proper}) by following the route of taking its history Poisson brackets with individual history variables. First, consider the brackets with the dynamical variables ${N}{\f}$ and ${\pf}$. (To get the field equations in their standard form, we need to turn ${\f}$ into a temporal density ${N}{\f}$; the momentum $\pf$ is a temporal density by itself.). It is easy to see that, by virtue of the spatial ultralocality of $\Theta$, the Liouville functional acting on ${N}{\f}$ and ${\pf}$ generates the temporal derivatives of the Dirac-ADM variables ${\f}$ and ${P_{\f}}$:
\begin{eqnarray}
\Big\{ \!\!\! \Big\{ \, ({\N}{\fa})(t,x) \, , \, \Theta  \, \Big\} \!\!\! \Big\} = { {\delta {\Theta}} \over {\delta P_A(t,x)} } = {\fa}_{,0}(t,x)    
\; , 
\label{eq:ADM equ1}
\\
\Big\{ \!\!\! \Big\{ \, {\pa}(t,x) \, , \, \Theta  \,  \Big\} \!\!\! \Big\} = {  {\delta {\Theta}} \over {\delta {\fa}(t,x)} } =  {P_A}_{,0}(t,x)   \; .
\label{eq:ADM equ2}
\end{eqnarray}
Notice that the derivatives (\ref{eq:ADM equ1})-(\ref{eq:ADM equ2}) are again temporal densities.

Next, and this is the key point, the history Poisson brackets of the dynamical variables ${N}{\f}$ and ${\pf}$ with the Hamiltonian functional ${\cal H}$, (\ref{eq:Hamiltonian Fnal}), turn out to be the same as the standard equal time Poisson brackets $\{ \, \, , \, \, \}$ of the variables ${\f}_{(t)}$ and ${P_{{\f}(t)}}$ with the {\it instantaneous Hamiltonian} 
\begin{equation}
H_{(t)0}[{\f}_{(t)},{P_{{\f}(t)}};Q_{(t)},N_{(t)},M_{(t)}] = \int dx \; \Big( {N}_{(t)0} \, H_{(t)} + {{M}_{(t)0}}^{\!\!\!1} \, H_{(t)1} \Big)(x) \; .
\label{eq:inst Hamiltonian Fnal}
\end{equation}
To state clearly what we mean requires some care. The supermomentum $H_1$  and super-Hamiltonian $H$, (\ref{eq:supermom})-(\ref{eq:superham}), are temporally ultralocal concomitants of the temporal scalars ${\f}$, $P_{\f}$ and $Q$. Being ultralocal, $H$ and $H_1$ are temporal scalars themselves. At a fixed moment $t \in \tman$, we smear $H_{(t)}$ and $H_{(t)1}$ by temporal densities $N_{(t)0}$ and ${{M}_{(t)0}}^{\!\!\!1}$ and obtain thereby the Hamiltonian (\ref{eq:inst Hamiltonian Fnal}). For a fixed $t \in \tman$, (\ref{eq:inst Hamiltonian Fnal}) is considered as a functional of purely spatial dynamical variables ${\f}_{(t)}$ and ${P_{{\f}(t)}}$ on which we impose the standard Poisson brackets
\begin{equation}
\{ \, {\Phi}_{(t)}{}^A(x)   \, , \,  P_{(t)B}(x')      \, \} = {\delta}^A_B \, {\delta}(x,x') \; .  
\label{eq:standard equal time}
\end{equation}
These brackets have the same form whatever moment $t \in \tman$ we choose and they are thus actually the brackets between maps
\begin{equation}
{\f}_{\xman} \, : \; \; {\xman} \; \rightarrow \; {\mman}  \; \; \; \; {\rm by} \; \; \; \; x \in {\xman} \; \mapsto \; {\phi} = {\f}_{\xman}(x) 
\in {\mman}
\label{eq:inst map1}
\end{equation}
and
\begin{equation}
{P}_{{\f}\xman} \, : \; \;  {\xman} \; \rightarrow \; {\rm T}^*{\mman}  \; \; \; \; {\rm by} \; \; \; \; x \in {\xman} \; \mapsto \; p = {P}_{{\f}\xman}(x) \in {\rm T}^*_{\f(x)}{\mman}
\label{eq:inst map2}
\end{equation}
which contain no reference to $\tman$. Let us emphasize that it makes no sense to speak about the bracket (\ref{eq:standard equal time}) between ${\fa}_{(t_1)}(x)$ at one moment $t_1 \in \tman$ and $P_{(t_2)B}(x')$ at another moment $t_2 \in \tman$, $t_2 \neq t_1$. All this being understood, let us re-express the history Poisson brackets $\Big\{ \!\!\! \Big\{ \, \, , \, \, \Big\} \!\!\! \Big\}$ of the dynamical history variables $N{\f}$ and $\pf$ with the Hamiltonian functional ${\cal H}$ in terms of the standard Poisson brackets $\{ \, \, , \, \, \}$ of the instantaneous variables ${\f}_{(t)}$ and ${P_{{\f}(t)}}$ with the {instantaneous Hamiltonian} $H_{(t)0}$:
\begin{eqnarray}
\Big\{ \!\!\! \Big\{ \, ({\N}{\fa})(t,x) \, , \, {\cal H} \, \Big\} \!\!\! \Big\} = {  {\delta {\cal H} } \over {\delta P_A(t,x)} } 
= \{ \, {\fa}_{(t)}(x) \, , \, H_{(t)0} \, \} \; ,
\nonumber
\\
\Big\{ \!\!\! \Big\{ \, {\pa}(t,x) \, , \, {\cal H}  \, \Big\} \!\!\! \Big\} = - {  {\delta {\cal H} } \over {\delta {\fa}(t,x)} } = 
\{ \, P_{(t)A}(x) \, , \, H_{(t)0}  \, \} \; .
\label{eq:Dirac-ADM equs}
\end{eqnarray}
In this sense, the history brackets $\Big\{ \!\!\! \Big\{ \, \, , \, \, \Big\} \!\!\! \Big\}$ of (\ref{eq:ind-bra1}) {\it induce} the timeless brackets $\{ \, \, , \, \, \}$ of (\ref{eq:standard equal time}) between the maps (\ref{eq:inst map1})-(\ref{eq:inst map2}). Let us stress that the validity of the statement (\ref{eq:Dirac-ADM equs}) does not depend on the detailed structure (\ref{eq:supermom}), (\ref{eq:superham}) of the supermomentum $H_1$ and super-Hamiltonian $H$, only on the fact that $H_1$ and $H$ are {\it temporally} ultralocal in their arguments.\footnote{Our particular $H$ and $H_1$ are also spatially ultralocal in $\pf$, or $P_{\f}$, but this plays no role in the derivation of (\ref{eq:Dirac-ADM equs}).} Putting the Liouville and Hamiltonian pieces together, we obtain the Dirac-ADM Hamilton equations
\begin{eqnarray}
{\Phi}_{(t)}{}^A{}_{,0}(x) - \{ \, {\Phi}^A_{(t)}(x) \, , \, H_{(t)0}  \, \} =
{  {\delta S } \over {\delta P_A(t,x)} } = \Big\{ \!\!\! \Big\{ \, ({\N}{\fa})(t,x) \, , \, S \, \Big\} \!\!\! \Big\} = 0 \; ,
\nonumber
\\
\!\!\!\!\!\!\!\! P_{(t)A,0}(x)  -  \{ \, P_{(t)A}(x) \, , \, H_{(t)0}  \, \} = 
- {  {\delta S } \over {\delta {\fa}(t,x)} } = \Big\{ \!\!\! \Big\{ \, {\pa}(t,x) \, , \, S  \, \Big\} \!\!\! \Big\} = 0 \; .
\nonumber
\\
\label{eq:final Dirac-ADM equs}
\end{eqnarray}

The remaining history Poisson brackets of the action with the momenta canonically conjugate to the geometric variables $N$, $M$ and $Q$ yield the constraints on the instantaneous data $\f$ and $P_{\f}$. The history bracket with $P_M$,
\begin{equation}
H_1(t,x) = {  {\delta {S}} \over {\delta {\M}(t,x)} } = \Big\{ \!\!\! \Big\{ \, {P}_1(t,x) \, , \, S \, \Big\} \!\!\! \Big\} = 0 \; , 
\label{eq:super mom con}
\end{equation}
gives directly the supermomentum constraint. The super-Hamiltonian constraint is obtained twice. The history bracket with ${{\Pi}_Q}$ gives it in an algebraically scaled version
\begin{equation}
H(t,x) = - \left( {{Q_{11}} \over \N} \right)(t,x) \, {  {\delta {S}} \over {\delta {Q_{11}}(t,x)} } = - \Big\{ \!\!\! \Big\{ \, \left( {{\Pi}}^{11} \,  {{Q_{11}} \over \N} \right) (t,x) \, , \, S \, \Big\} \!\!\! \Big\}  = 0 \; , 
\label{eq:super Ham con1}
\end{equation}
while the history bracket with ${P_N}$ gives it only modulo the first set of Hamilton's equations. This happens because $N$ enters $S$ not only as the smearing function, but also through the momentum $P_{\f} = {N}^{-1} {\pf}$. We thus get
\begin{equation}
H(t,x) = {  {\delta {S}[{P_{\f}}]} \over {\delta {\N}(t,x)} } = \Big\{ \!\!\! \Big\{ \, {P}(t,x) \, , \, S \, \Big\} \!\!\! \Big\} +
P_A(t,x) \, \Big\{ \!\!\! \Big\{ \, {\fa}(t,x) \, , \, S \, \Big\} \!\!\! \Big\} = 0 \; .
\label{eq:super Ham con2}
\end{equation}
This redundancy in obtaining the Hamiltonian constraint is, of course, related to the conformal invariance of the Polyakov action which implies that the energy-momentum tensor $T^{\mu\nu}$ in (\ref{eq:ST-cons}) is necessarily tracefree.

This completes our argument that the instantaneous Dirac-ADM canonical formalism follows from the canonical history action.

\section{Diffeomorphisms of $\yman$, $\tman$ and $\xman$}

To repeat, our description of the canonical formalism for the string model is based on three manifolds, the spacetime $\yman$, time $\tman$, and space $\xman$. The elements $y$ of $\yman$ are called {\it events}, the elements $t$ of $\tman$ are called {\it moments}, and the elements $x$ of $\xman$ are called {\it points}. Each of these manifolds is subject to its own group of diffeomorphisms. Spacetime diffeomorphisms ${\sf D} \in \rm{Diff}{\yman}$ move the events in $\yman$ but keep the moments $t$ and points $x$ fixed, temporal diffeomorphisms ${D_{\tman}} \in {\rm Diff}{\tman}$ move the moments $t$ in $\tman$ but keep the events $y$ and the points $x$ fixed, and finally spatial diffeomorphisms ${D_{\xman}} \in {\rm Diff}{\xman}$ move the points $x$ in $\xman$ but keep the events $y$ and moments $t$ fixed. We want to show that our formalism is invariant under all of these diffeomorphisms and that each of them can be implemented by a symplectomorphism in the history phase space ${\cal G}$.

We start by studying how these diffeomorphisms act on the time map $\T$ and space map $\sf X$. A spacetime diffeomorphism 
\begin{equation}
{\sf D} \in \rm{Diff}{\yman} : {\yman} {\rightarrow} {\yman} \; \; \rm{by} \; \; y' \in {\yman} \; \; {\mapsto} \; \; y={\sf D}(y') \in {\yman} 
\label{eq:diff}
\end{equation}
moves the points $y$ in $\yman$ and thereby acts on the maps $\T$ and $\sf X$ by the pullbacks
\begin{eqnarray}
{\T}' &=& {\sf D}^* {\T} = {\T} \circ {\sf D} \; , 
\label{eq:acts on}
\\
{\sf X}'&=& {\sf D}^* {\sf X} = {\sf X} \circ {\sf D} \; .
\label{eq:acts on2}
\end{eqnarray}
This implies that the mapping $Y$ of $\tman \times \xman$ onto $\yman$ is, in accordance with (\ref{eq:acts on})-(\ref{eq:acts on2}), pushed forward by ${\sf D}$ into
\begin{equation}
{Y}' = {\sf D}_* {Y} = {\sf D^{-1}} \circ  {Y}  \; . 
\label{eq:acts on Y}
\end{equation}

While ${\sf D}$ does not move the moments $t$ in $\tman$, it changes the time map $\T$ and hence the time foliation ${\Sigma}$ in $\yman$. It sends the instant 
\begin{equation}
{\Sigma}_{(t)} = {\Sigma}^{\T}_{(t)} = \Big\{ y {\in} {\yman} : {\T}(y) = t {\in} {\tman} \Big\} 
\label{eq:sigmaT}
\end{equation}
of the original time foliation ${\Sigma}$ into an instant 
\begin{equation}
{\Sigma}^{'}_{(t)} = {\Sigma}^{{\T}'}_{(t)} = \Big\{ y' {\in} {\yman} : {\T}'(y') = {\T} \circ {\sf D} (y') = t {\in} {\tman} \Big\} 
\label{eq:sigmaT'}
\end{equation}
of a different time foliation ${\Sigma}'$: In general, the instant ${\Sigma}^{'}_{(t)}$ is not only different from the instant ${\Sigma}_{(t)}$, but also from all the other instants ${\Sigma}_{(t')}$, $t' \in {\tman}$, of the original time foliation ${\Sigma}$.

Similarly, while ${\sf D}$ does not move the points $x$ in $\xman$, it changes the space map $\sf X$ and hence the reference frame $C$ in $\yman$. It sends the reference worldline 
\begin{equation}
C_{(x)} = C^{\sf X}_{(x)} = \Big\{ y {\in} {\yman} : {\sf X}(y) = x {\in} {\xman} \Big\} 
\label{eq:CX}
\end{equation}
of the original reference frame $C$ onto a reference worldline
\begin{eqnarray}
C^{'}_{(x)} = {C}^{{\sf X}'}_{(x)} &=& \Big\{ y' {\in} {\yman} : {\sf X}'(y') = {\sf X} \circ {\sf D} (y') = x {\in} {\xman} \Big\} 
\label{eq:CX'}
\end{eqnarray}
of a different reference frame $C'$: In general, the reference worldline ${C}^{'}_{(x)}$ is not only different from the reference worldline $C_{(x)}$, but also from all the other reference worldlines ${C}_{(x')}$, $x' \in {\xman}$, of the original reference frame ${C}$.

A temporal diffeomorphism 
\begin{equation}
{D_{\tman}} \in {\rm Diff}{\tman} : {\tman} {\rightarrow} {\tman} \; \; {\rm by} \; \; t' \in {\tman} \; \; {\mapsto} \; \; t={D_{\tman}}(t') \in {\tman}
\label{eq:t-diff}
\end{equation}
moves the moments $t$ in $\tman$ and thereby acts on the time map $\T$ by the pushforward
\begin{equation}
{\T}' = {D_{{\tman}*} {\T}} = {D_{\tman}}^{-1} \circ {\T} \; .
\label{eq:t-acts on}
\end{equation}
It sends the instant (\ref{eq:sigmaT}) into an instant  
\begin{eqnarray}
{\Sigma}^{'}_{(t')} = {\Sigma}^{{\T}'}_{(t')} &=& \Big\{ y {\in} {\yman} : {\T}'(y) = {D_{\tman}}^{-1} \circ {\T}(y) = t' {\in} {\tman} \Big\} 
\nonumber
\\
&=& \Big\{ y {\in} {\yman} : {\T}(y) = {D_{\tman}}(t') = t {\in} {\tman} \Big\} = {\Sigma}^{\T}_{(t)} \in {\Sigma}
\label{eq:sigma'T'2}
\end{eqnarray}
of the original time foliation. In other words, the foliation ${\Sigma}^{\T}$ depends only on the equivalence class $\big\{ {\T} \big\}$ of time maps modulo temporal diffeomorphisms. Of course, temporal diffeomorphisms ${D_{\tman}} \in {\rm Diff}{\tman}$ leave the space map $\sf X$, and hence not only the reference frame $C$ but also its individual reference worldlines $C_{(x)}$, fixed. 
The mapping $Y$ of $\tman \times \xman$ onto $\yman$ is pulled back by ${D_{\tman}} \in {\rm Diff}{\tman}$ into 
\begin{equation}
Y' =  Y  \circ \Big( \, {D_{\tman}} \times {\rm Id}_{\xman} \, \Big) \; .
\end{equation}

Spatial diffeomorphisms can be handled in the same manner as the temporal ones.
A spatial diffeomorphism
\begin{equation}
{D_{\xman}} \in {\rm Diff}{\xman} : {\xman} {\rightarrow} {\xman} \; \; {\rm by} \; \; x' \in {\xman} \; \; {\mapsto} \; \; x={D_{\xman}}(x') \in {\xman}
\label{eq:x-diff}
\end{equation}
moves the points $x$ in $\xman$ and thereby acts on the space map $\sf X$ by the pushforward
\begin{equation}
{\sf X}' = {D_{{\xman}*} {\sf X}} = {D_{\xman}}^{-1} \circ {\sf X} \; .
\label{eq:x-acts on}
\end{equation}
It sends the reference worldline (\ref{eq:CX}) to the reference worldline  
\begin{eqnarray}
{C}^{'}_{(x')} = {C}^{{\sf X}'}_{(x')} &=& \Big\{ y {\in} {\yman} : {\sf X}'(y) = {D_{\xman}}^{-1} \circ {\sf X}(y) = x' {\in} {\xman} \Big\} 
\nonumber
\\
&=& \Big\{ y {\in} {\yman} : {\sf X}(y) = {D_{\xman}}(x') = x {\in} {\xman} \Big\} = {C}^{\sf X}_{(x)} \in C
\label{eq:C'X'2}
\end{eqnarray}
of the original reference frame. In other words, the reference frame ${C}^{\sf X}$ depends only on the equivalence class $\big\{ {\sf X} \big\}$ of space maps modulo spatial diffeomorphisms. Spatial diffeomorphisms ${D_{\xman}} \in {\rm Diff}{\xman}$ leave the time map $\T$, and hence not only the time foliation ${\Sigma}$ but also its individual instants ${\Sigma}_{(t)}$, fixed. 
The mapping $Y$ of $\tman \times \xman$ onto $\yman$ is pulled back by ${D_{\xman}} \in {\rm Diff}{\xman}$ into 
\begin{equation}
Y' =  Y  \circ \Big( \, {\rm Id}_{\tman} \times {D_{\xman}} \, \Big) \; .
\end{equation}

In the rest of this paper, we shall limit our attention to infinitesimal diffeomorphisms which are the elements ${\sf U} \in {\rm diff}{\yman}$, $V \in {\rm diff}{\tman}$ and $W \in {\rm diff}{\xman}$ of the Lie algebras of the diffeomorphism groups ${\rm Diff}{\yman}$, ${\rm Diff}{\tman}$ and ${\rm Diff}{\xman}$. The Lie bracket $[ \, \, \, \, ]$ on those algebras differs by sign from the Lie bracket $[ \, , \, ]$ of their elements considered as vector fields ${\sf U}^{\alpha}$ on $\yman$, 
$V^{0}$ on $\tman$ and $W^{1}$ on $\xman$ which is defined as
\begin{eqnarray}
\!\!\!\! && \!\!\!\! [ \, {\sf U}_{(1)}(y) \, , \, {\sf U}_{(2)}(y) \, ]^{\alpha} = \Big( {{\sf U}_{(1)}}^{\beta} \, {\partial}_{\beta}{{\sf U}_{(2)}}^{\alpha} -   {{\sf U}_{(2)}}^{\beta} \,  {\partial}_{\beta}{{\sf U}_{(1)}}^{\alpha} \Big)(y) \; , 
\label{eq:definition-Lie1}
\\
\!\!\!\! && \!\!\!\! {[ \, {V}_{(1)}(t) \,  , \,  {V}_{(2)}(t) \,  ]}^{0} = \Big( {{V}_{(1)}}^{0} \,  {\partial}_{0}{{V}_{(2)}}^{0} -  {{V}_{(2)}}^{0} \,  {\partial}_{0}{{V}_{(1)}}^{0} \Big)(t) 
\label{eq:definition-Lie2} 
\\
\!\!\!\!\!\!\!\!\!\!\!\!\!\!\!\! {\rm and} \!\!\!\! &&  
\nonumber
\\
\!\!\!\! && \!\!\!\! {[ \, {W}_{(1)}(x) \, , \, {W}_{(2)}(x) \, ]}^{1} = \Big( {{W}_{(1)}}^{1} \,  {\partial}_{1}{{W}_{(2)}}^{1} -  {{W}_{(2)}}^{1} \, {\partial}_{1}{{W}_{(1)}}^{1} \Big)(x)   \; .
\label{eq:definition-Lie3}
\end{eqnarray}
The vector field ${\sf U}^{\alpha}$ acts on the time map ${\T}$ and the space map ${\sf X}$ by
\begin{eqnarray}
&& {\sf U}({\T}^0)(y) = {\cal L}_{({\sf U})}{\T}^0(y) = \Big( {{\T}^0}_{,\alpha}    \, {\sf U}^{\alpha} \Big)(y) \; ,
\nonumber
\\
&& {\sf U}({\sf X}^1)(y) = {\cal L}_{({\sf U})}{\sf X}^1(y) = \Big( {{\sf X}^1}_{,\alpha} \, {\sf U}^{\alpha} \Big)(y) \; ,
\label{eq:U on TX}
\end{eqnarray} 
the vector field $V^{0}(t)$ acts on these maps by
\begin{eqnarray}
{V}({\T}^0)(y) &=& - {V}^0({\T}(y)) \; ,
\nonumber
\\
{V}({\sf X}^1)(y) &=& 0 \; ,
\label{eq:V on TX}
\end{eqnarray} 
and the vector field $W^{1}(x)$ acts on them by
\begin{eqnarray}
{W}({\T}^0)(y) &=& 0 \; ,
\nonumber
\\
{W}({\sf X}^1)(y) &=& - {W}^1({\sf X}(y)) \; .
\label{eq:W on TX}
\end{eqnarray} 
The difference between the actions (\ref{eq:V on TX}), (\ref{eq:W on TX}) of ${\rm diff}{\tman}$ and ${\rm diff}{\xman}$ and the action (\ref{eq:U on TX}) of ${\rm diff}{\yman}$ reflects the fact that temporal and spatial diffeomorphisms act on the ranges rather than on the domain of the maps $\T$ and $\sf X$.

The map $Y$ is the inverse map of ${\T} \times {\sf X}$ and hence it changes under infinitesimal diffeomorphisms in exactly the opposite way: $Y(t,x)[{\T},{\sf X}]$ is changed by the Lie derivative when we act on it by $V \in {\rm diff}{\tman}$,
\begin{equation}
{V}(Y^{\alpha})(t,x) = {\cal L}_{(V)} Y^{\alpha}(t,x) = {V}^{0}(t) \, \, {\partial}_{0}{Y^{\alpha}}(t,x)  \; ,
\label{eq:V on Y}
\end{equation}
or when we act on it by $W \in {\rm diff}{\xman}$,
\begin{equation}
{W}(Y^{\alpha})(t,x) = {\cal L}_{(W)} Y^{\alpha}(t,x) = {W}^{1}(x) \, \, {\partial}_{1}{Y^{\alpha}}(t,x)  \; ,
\label{eq:W on Y}
\end{equation}
while it is changed by
\begin{equation}
{\sf U}(Y^{\alpha})(t,x) = - {\sf U}^{\alpha}(Y(t,x)) 
\label{eq:U on Y}
\end{equation}
when we act on it by ${\sf U} \in {\rm diff}{\yman}$. 

Next, let us consider the action of infinitesimal diffeomorphisms on the spacetime metric $\G$ and its alternative description by $Q$ and ${N}$, ${M}$ on $\tman \times \xman$. A spacetime vector field ${\sf U} \in {\rm diff}{\yman}$ acts on the spacetime metric $\G$ by the Lie derivative,
\begin{eqnarray}
{\sf U}({\Gab})(y) &=& {\cal L}_{({\sf U})}{\Gab}(y) 
\nonumber
\\
&=& \Big(  {\sf U}^{\gamma} \, {\partial}_{\gamma} \, {\Gab} +  {\sf G}_{\gamma\beta} \, {\partial}_{\alpha} \, {\sf U}^{\gamma} + {\sf G}_{\alpha\gamma} \, {\partial}_{\beta} \, {\sf U}^{\gamma}  \Big)(y) \; , 
\label{eq:U on G}
\end{eqnarray}
while the vector fields $V \in {\rm diff}{\tman}$ and $W \in {\rm diff}{\xman}$ annihilate it,
\begin{equation}
{V}({\Gab})(y) = 0 \; \; , \; \; {W}({\Gab})(y) = 0 \; .
\label{eq:VW on G}
\end{equation} 
On the other hand, from the dependence of $Q$, ${N}$ and ${M}$ on the spacetime variables $\G$, $\T$ and $\X$, (\ref{eq:final expressions}), we get that ${\sf U} \in {\rm diff}{\yman}$ annihilates them,
\begin{equation}
{\sf U}({\qxx})(t,x) = 0 \; \; \; \; , \; \; \; \; {\sf U}({\N})(t,x) = 0 = {\sf U}({\M})(t,x) \; ,
\label{eq:U on gNM}
\end{equation} 
and the vector fields $V \in {\rm diff}{\tman}$ and $W \in {\rm diff}{\xman}$ act on them by the Lie derivatives: 
\begin{eqnarray}
&& {V}({\qxx})(t,x) = {\cal L}_{(V)}{\qxx}(t,x) = {V}^0(t) \, {\partial}_0  {\qxx}(t,x) \; ,
\nonumber
\\
&& {V}({\N})(t,x) = {\cal L}_{(V)}{\N}(t,x) = {\partial}_0  \Big( {V}^0(t) \, {\N}(t,x) \Big) \; ,
\nonumber
\\
&& {V}({\M})(t,x) = {\cal L}_{(V)}{\M}(t,x) = {\partial}_0  \Big( {V}^0(t) \, {\M}(t,x) \Big) \; ,
\label{eq:V on gNM}
\\
\nonumber
\\
&& {W}({\qxx})(t,x)  = {\cal L}_{(W)}{\qxx}(t,x) = {W}^1(x) \,  {\partial}_1  {\qxx}(t,x) +  2 {\qxx}(t,x) \,  {\partial}_1 {W}^1(x) \; , \;
\nonumber
\\
&& {W}({\N})(t,x) = {\cal L}_{(W)}{\N}(t,x) = {W}^1(x) \,  {\partial}_1  {\N}(t,x)  \; ,
\nonumber
\\
&& {W}({\M})(t,x) = {\cal L}_{(W)}{\M}(t,x) = {W}^1(x) \,  {\partial}_1 {\M}(t,x)  -  {\M}(t,x)  \, {\partial}_1  {W}^1(x) \; .
\nonumber
\\
\label{eq:W on gNM}
\end{eqnarray} 
The action of infinitesimal diffeomorphisms on the geometric momenta ${\PG}$ or their split version $P_N$, ${P_M}$ and ${{\Pi}_H}$ follows the same pattern and we shall not write the corresponding equations.

Finally, we turn our attention to the canonical variables $\F$, $\PF$ and their counterparts $\f$, ${\pf}$. Under spacetime diffeomorphisms, $\F$ behaves as a scalar and $\PF$ as a scalar density:
\begin{eqnarray}
{\sf U}({\FA})(y) \! \! \! \! &=& \! \! \! \! {\cal L}_{({\sf U})}{\FA}(y) = \Big( {\sf U}^{\alpha} \, {\partial}_{\alpha}{\FA} \Big)(y)  \; ,
\nonumber
\\
{\sf U}({\PA})(y) \! \! \! \! &=& \! \! \! \! {\cal L}_{({\sf U})}{\PA}(y) = {\partial}_{\alpha} \Big( {\sf U}^{\alpha} \, {\PA} \Big)(y) \; .
\label{eq:U on FP}
\end{eqnarray} 
The spacetime variables $\F$, $\PF$ are annihilated by the action of the vector fields $V \in {\rm diff}{\tman}$ and $W \in {\rm diff}{\xman}$,
\begin{equation}
{V}({\FA})(y) = 0 = {V}({\PA})(y) \; \; \; \; ,  \; \; \; \; {W}({\FA})(y) = 0 = {W}({\PA})(y) \; ,
\label{eq:VW on FP}
\end{equation} 
and the variables $\f$, $\pf$ are annihilated by the action of ${\sf U} \in {\rm diff}{\yman}$, 
\begin{equation}
{\sf U}({\fa})(t,x) = 0 = {\sf U}({\pa})(t,x)  \; .
\label{eq:U on fp}
\end{equation} 
The action of $V \in {\rm diff}{\tman}$ and $W \in {\rm diff}{\xman}$ on $\f$ and $\pf$ follows the pattern with the Lie derivative,
\begin{eqnarray}
\!\!\!\!\!\!\!\! && {V}({\fa})(t,x) = {\cal L}_{(V)}{\fa}(t,x) = {V}^0(t) \, {\partial}_0{\fa}(t,x) \; ,
\nonumber
\\
\!\!\!\!\!\!\!\! && {V}({\pa})(t,x) = {\cal L}_{(V)}{\pa}(t,x) = {\partial}_0 \Big( {V}^0(t) \, {\pa}(t,x)  \Big) \; ,
\label{eq:V on fp}
\end{eqnarray} 
and
\begin{eqnarray}
\!\!\!\!\!\!\!\! && {W}({\fa})(t,x) = {\cal L}_{(W)}{\fa}(t,x) = {W}^1(x) \, {\partial}_1{\fa}(t,x) 
\; ,
\nonumber
\\
\!\!\!\!\!\!\!\! && {W}({\pa})(t,x) = {\cal L}_{(W)}{\pa}(t,x) = {\partial}_1 \Big( {W}^1(x) \, {\pa}(t,x) \Big)  \; .
\label{eq:W on fp}
\end{eqnarray} 
This concludes our little pedantic enumeration of how infinitesimal diffeomorphisms affect the history variables.

\section{Diffeomorphism Invariance of the Action}

We are now ready to show that all forms of the action $S$, the Polyakov action (\ref{eq:ST-L}), the canonical history action (\ref{eq:ST-H}), and the Dirac-ADM action (\ref{eq:ADM proper})-(\ref{eq:Hamiltonian Fnal}), (\ref{eq:supermom})-(\ref{eq:superham}), are invariant under all diffeomorphisms $\sf D$, $D_{\tman}$ and $D_{\xman}$ of the manifolds $\yman$, $\tman$ and $\xman$. This statement is valid for the action written as an integral over all of $\yman$ or $\zman$ and has nothing to do with the domains $\cal D$ of variation. The invariance of the action thus holds on the space of all virtual histories, prior to fixing the ends of the field variables on $\cal B$ for the purpose of deriving the field equations. Of course, the action is in particular invariant when evaluated on an actual history. 

Let us write the spacetime action as a functional 
\begin{equation}
S[{\sf \Gamma}] = {\int}_{\yman} d^2y \, {\sf L}(y)[{\sf \Gamma}]
\label{eq:as a functional}
\end{equation}
of the canonical fields ${\sf \Gamma}$ in $\cal G$, (\ref{eq:ext phase space}). Here, the Lagrangian ${\sf L}[{\sf \Gamma}]$ is a concomitant of the fields containing their derivatives up to the first order. The action does not need to depend on all the canonical variables: the canonical history action (\ref{eq:ST-H}) does not depend on $\X$, $\PZ$ and $\PG$, and the Polyakov action (\ref{eq:ST-L}) does also not depend on $\T$. The missing variables simply do not matter in our consideration.

Let us see how the infinitesimal diffeomorphisms ${\sf U} \in {\rm diff}{\yman}$, ${V} \in {\rm diff}{\tman}$ and ${W} \in {\rm diff}{\xman}$ affect the action. We assume that the vector fields ${\sf U}$, $V$ and $W$ have compact supports ${\cal U} \in \yman$, ${\cal V} \in \tman$ and ${\cal W} \in \xman$. These may be arbitrarily large and have no connection whatsoever with the domains ${\cal D}$ of variation. We have itemized in (\ref{eq:U on TX}), (\ref{eq:U on G}) and (\ref{eq:U on FP}) how the infinitesimal diffeomorphisms act on the tensor fields from which the Lagrangian ${\sf L}$ is constructed. This induces the action of ${\sf U}$ on the Lagrangian $\sf L$:
\begin{equation}
{\sf U}({\sf L})(y)[{\sf \Gamma}] =  {\int}_{\yman} d^2y' \, { {\delta {\sf L}(y)[{\sf \Gamma}]  }  \over {\delta {\sf \Gamma}(y')}  }  \, {\cal L}_{\sf U}{\sf \Gamma}(y')  \; ,
\label{eq:on the Lagrangian}
\end{equation}
where the summation over all ${\sf \Gamma}$'s is implied on the right-hand side of (\ref{eq:on the Lagrangian}).  
An infinitesimal diffeomorphism ${\sf U} \in {\rm diff}{\yman}$ then acts on the action functional (\ref{eq:as a functional}) by
\begin{equation}
{\sf U}(S)[{\sf \Gamma}] = S[{\sf U}({\sf \Gamma})] =  {\int}_{\yman} d^2y \,   {\sf U}({\sf L})(y)[{\sf \Gamma}]  \; .
\label{eq:mpla-mpla}
\end{equation}

The punch line of the argument is simple. One can see by direct inspection that the Lagrangians of the Polyakov action (\ref{eq:ST-L}) and of the canonical history action (\ref{eq:ST-H}) are {\it scalar density} concomitants of their arguments ${\sf \Gamma}$. Therefore, one can easily verify that (\ref{eq:on the Lagrangian}) yields the result 
\begin{equation}
{\sf U}({\sf L})(y)[{\sf \Gamma}] =  {\cal L}_{\sf U}{\sf L}(y)[{\sf \Gamma}] = {\partial}_{\alpha} \Big(  {\sf U}^{\alpha}(y) \, \, {\sf L}(y)[{\sf \Gamma}]   \Big)    
\label{eq:yields the result}
\end{equation}
which one expects for a scalar density. 
Equation (\ref{eq:mpla-mpla}) then reduces by the Gauss theorem to the identity
\begin{equation}
{\sf U}(S)[{\sf \Gamma}] =  {\int}_{\yman} d^2y \,  {\partial}_{\alpha} \Big( {\sf U}^{\alpha}(y) \, {\sf L}(y)[{\sf \Gamma}] \Big) \equiv 0 
\label{eq:final Gauss thm}
\end{equation}
because ${\sf U}$ vanishes on $\yman$ outside a compact region ${\cal U}$.
The {\it identity} (\ref{eq:final Gauss thm}) means that the actions (\ref{eq:ST-L}) and (\ref{eq:ST-H}) are invariant under all infinitesimal spacetime diffeomorphisms with compact support (or vanishing sufficiently fast at infinity).

The invariance of actions (\ref{eq:ST-L}) and (\ref{eq:ST-H}) under temporal or spatial diffeomorphisms, $V \in {\rm diff}{\tman}$ or $W \in {\rm diff}{\xman}$, is quite obvious. The dynamical variables $\F$ and $\PF$ and the metric $\G$ carry no reference to $\tman$ or $\xman$, the $\X$ mapping (which changes under ${\rm Diff}\xman$) is absent from both actions, and the $\T$ mapping (which changes under ${\rm Diff}\tman$) is absent from the Polyakov action (\ref{eq:ST-L}) while it appears in the canonical history action (\ref{eq:ST-H}) only through the vector field $\sf E$ which, according to (\ref{eq:e-field}), is manifestly invariant under infinitesimal temporal diffeomorphisms. Formally, $\F$, $\PF$ and $\G$ are annihilated by $V$ or $W$, (\ref{eq:VW on FP}) and (\ref{eq:VW on G}), and 
\begin{equation}
V({\sf E}^{\alpha})(y) = 0 
\label{eq:V on E}
\end{equation}
by virtue of (\ref{eq:e-field}). Therefore, 
\begin{equation}
V(S)[{\sf {\Gamma}}] \equiv 0 \; \; \; \; , \; \; \; \; W(S)[{\sf \Gamma}] \equiv 0 
\label{eq:V on S}
\end{equation}
for both functionals (\ref{eq:ST-L}) and (\ref{eq:ST-H}) of ${\sf \Gamma}$.

As long as the history action (\ref{eq:ADM proper})-(\ref{eq:Hamiltonian Fnal}) is considered as a functional of ${\sf \Gamma}$, as in (\ref{eq:ADM action2}), it coincides with the action (\ref{eq:ST-H}) and shares its invariances. 
When we consider it as a functional of the history variables ${{\Gamma}}$ on $\zman$, we can rewrite it as
\begin{equation}
S[{\Gamma}] = {\int}_{\tman} dt^0 {\int}_{\xman} dx  \, {L}(t,x)[{\Gamma}]  
\label{eq:in the form}
\end{equation}
and study its behavior under infinitesimal diffeomorphisms $\sf U$, $V$ and $W$ directly. The variables ${\Gamma}$ do not change under ${\sf U} \in {\rm diff}{\yman}$ and hence 
\begin{equation}
U(S)[{\Gamma}] \equiv 0 \; .
\label{eq:U on Sz}
\end{equation}

An infinitesimal diffeomorphism $V \in {\rm diff}{\tman}$ acts on the Lagrangian $L[{\Gamma}]$ by
\begin{equation}
{V}({L})(t,x)[{\Gamma}] = {\int}_{\zman} d^2z' \, { {\delta {L}(t,x)[{\Gamma}]  }  \over {\delta {\Gamma}(z')}  }  \, {\cal L}_{V}{\Gamma}(z')  
\label{eq:and the Lagrangian by}
\end{equation}
and through it on the action:
\begin{equation}
{V}(S)[{\Gamma}] = {\int}_{\tman} dt^0 {\int}_{\xman} dx \, {V}({L})(t,x)[{\Gamma}]  \; .
\label{eq:fundamental formula2}
\end{equation}
An analogous formula holds for spatial diffeomorphisms ${W} \in {\rm diff}{\xman}$. Equations (\ref{eq:V on gNM})-(\ref{eq:W on gNM}) and (\ref{eq:V on fp})-(\ref{eq:W on fp}) give the Lie derivatives ${\cal L}_V{\Gamma}$ and ${\cal L}_W{\Gamma}$, and the inspection of the action (\ref{eq:ADM proper})-(\ref{eq:Hamiltonian Fnal}), (\ref{eq:supermom})-(\ref{eq:superham}) reveals that $L$ is a scalar density concomitant of $\Gamma$ both in the temporal and the spatial arguments $t$ and $x$. 
Therefore
\begin{eqnarray}
V(L)(t,x) = {\partial}_0 \Big( V^0(t) \, L(t,x)   \Big) \; ,
\nonumber
\\
W(L)(t,x) = {\partial}_1 \Big( W^1(x) \, L(t,x)   \Big) \; .
\label{eq:fratelo}
\end{eqnarray}
This, exactly as in the case of spacetime diffeomorphisms $\sf U$ acting on $S[{\sf \Gamma}]$, allows us to handle the fundamental formula (\ref{eq:fundamental formula2}) by the Gauss theorem and conclude that
\begin{equation}
{V}(S)[{\Gamma}] \equiv 0 \; \; \; \; , \; \; \; \;  {W}(S)[{\Gamma}] \equiv 0     
\label{eq:VW on S}
\end{equation}
for all vector fields $V$ and $W$ with compact supports $\cal V$ and $\cal W$ on $\tman$ or $\xman$. The identities (\ref{eq:U on Sz}) and (\ref{eq:VW on S}) express the invariance of the Dirac-ADM action (\ref{eq:ADM proper})-(\ref{eq:Hamiltonian Fnal}) under infinitesimal diffeomorphisms ${\sf U} \in {\rm diff}{\yman}$, ${V} \in {\rm diff}{\tman}$ and ${W} \in {\rm diff}{\xman}$.

\section{Diffeomorphisms as Symplectomorphisms}

We are now able to conclude our treatment of the string model on the history phase space $\cal G$ by showing that the symplectic structure of $\cal G$  allows us to represent all diffeomorphisms ${\sf U} \in {\rm diff}{\yman}$, ${V} \in {\rm diff}{\tman}$ and ${W} \in {\rm diff}{\xman}$ by infinitesimal symplectomorphisms on $\cal G$. Infinitesimal symplectomorphisms of $\cal G$ are generated by Poisson brackets of the variables ${\sf \Gamma}(y)$ with functionals $F \in {\cal F}({\cal G})$. Let us consider first ${\rm diff}{\yman}$. Our aim is to map vector fields ${\sf U}$ on $\yman$ to functionals ${\sf R}(U)[{\sf \Gamma}]$ on $\cal G$,
\begin{equation}
{\sf R} \; \; : \; \; {\rm diff}{\yman} \; \; \rightarrow \; \;  {\cal F}[{\cal G}]  \; \; {\rm by}  \; \;  {\sf U} \in {\rm diff}{\yman} 
\; \; \mapsto \; \; {\sf R}({\sf U})[{\sf \Gamma}] \in {\cal F}[{\cal G}] \; ,
\label{eq:aarep'}
\end{equation} 
so that the history Poisson bracket of ${\sf R}({\sf U})[{\sf \Gamma}]$ with the fundamental canonical variables ${\sf \Gamma}$ reproduces their Lie derivatives along the vector field  ${\sf U}$:
\begin{equation}
{\cal L}_{({\sf U})}{\sf \Gamma}(y) = \Big\{ \!\!\! \Big\{ \, {\sf \Gamma}(y)  \, , \, {\sf R}({\sf U})[{\sf \Gamma}]  \, \Big\} \!\!\! \Big\} \; .
\label{eq:aaEq1}
\end{equation} 
More generally, if $F[{\sf \Gamma}]$ is an arbitrary tensor concomitant of the canonical variables, the relation
\begin{equation}
\Big\{ \!\!\! \Big\{ \, F(y)[{\sf \Gamma}]   \, , \,  {\sf R}({\sf U})[{\sf \Gamma}]  \, \Big\} \!\!\! \Big\} = {\cal L}_{({\sf U})}F(y)[{\sf \Gamma}] 
\label{eq:cov equs}
\end{equation}
will hold as a consequence of the action (\ref{eq:aaEq1}) of ${\sf R}({\sf U})[{\sf \Gamma}]$ on the individual variables. 
All field equations (\ref{eq:variation}) have the form of tensor equations $F[{\sf \Gamma}] = 0$ and hence they will transform covariantly, as in (\ref{eq:cov equs}), under infinitesimal canonical transformations generated by ${\sf R}({\sf U})[{\sf \Gamma}]$.

For ${\sf U} \in {\rm diff}{\yman}$, the map (\ref{eq:aarep'}) is given by
\begin{equation}
{\sf R}{({\sf U})}[{\sf \Gamma}] = {\int}_{\yman} d^2y \; \Big( {\PA} \, {\cal L}_{({\sf U})} {\FA}  + {\Pab} \, {\cal L}_{({\sf U})} {\Gab} + {\Pa} \, {\cal L}_{({\sf U})} {\Za} \Big)(y) \; .
\label{eq:aarep}
\end{equation} 
Notice that the integral in (\ref{eq:aarep}) is taken over the whole spacetime $\yman$. For (\ref{eq:aarep}) to be well defined, we work with vector fields ${\sf U}$ with compact support ${\cal U}$. It is easy to check that the functional (\ref{eq:aarep}) reaches its goal (\ref{eq:aaEq1}) anywhere by virtue of the fundamental Poisson brackets (\ref{eq:ST-bra1})-(\ref{eq:ST-bra3}) and (\ref{eq:ST-bra4}).

It is easy to check that for two vector fields ${\sf U}_{(1)}$ and ${\sf U}_{(2)}$ on ${\cal U}$ we have
\begin{equation}
\Big\{ \!\!\! \Big\{ \, {\sf R}{({\sf U}_{(1)})}      \, , \,  {\sf R}{({\sf U}_{(2)})}       \, \Big\} \!\!\! \Big\} = {\sf R}{( [ \, {\sf U}_{(1)} \, , \,  {\sf U}_{(2)}       \, ] )} \; .
\label{eq:aaalgebra}
\end{equation} 
This shows that the Lie algebra ${\rm diff}{\yman}$ (whose bracket $[ \, {\sf U}_{(1)} \, \,  {\sf U}_{(2)} \, ]$ differs by sign from the Lie bracket $[ \, {\sf U}_{(1)} \, , \,  {\sf U}_{(2)} \, ]$ of the elements ${\sf U} \in  {\rm diff}{\yman}$ considered as vector fields on $\yman$) is antihomomorphically {represented} by the Poisson algebra of functionals ${\sf R}{({\sf U})}[{\sf \Gamma}]$ on ${\cal G}$, i.e., by infinitesimal symplectomorphisms. Our use of the symbol ${\sf R}$ emphasizes that the mapping (\ref{eq:aarep'}), (\ref{eq:aarep}) is a representation.

The canonical transformation (\ref{eq:transf1a})-(\ref{eq:transf4}) allows us to write the generator (\ref{eq:aarep}) as a functional of the new canonical variables $\bar{\Z}$, ${\bar {\sf \Pi}}_{\bar {\Z}}$ and ${\Gamma}$ on ${\cal G}$.
We find that the generator 
\begin{equation}
{\sf R}{({\sf U})}[{\bar {\Z}}, {\bar {\sf \Pi}}_{\bar {\Z}}] = {\int}_{\yman} d^2y \; \Big( {\bar{\sf \Pi}}_a \, {\cal L}_{({\sf U})}{\bar{\Z}}^a \Big)(y) \; .
\label{eq:aarep2}
\end{equation} 
depends only on ${\bar {\Z}}$ and ${\bar {\sf \Pi}}_{\bar {\Z}}$, not on the variables ${\Gamma}$. It correctly reproduces the Lie derivatives along ${\sf U}$ of the new fundamental canonical variables ${\bar {\Z}}$, ${\bar {\sf \Pi}}_{\bar {\Z}}$ and ${\Gamma}$ and hence also the Lie derivative of an arbitrary tensor concomitant of them. It is also not difficult to check that the Poisson bracket action of ${\sf R}({\sf U})[{\bar {\Z}}, {\bar {\sf \Pi}}_{\bar {\Z}}]$ on an arbitrary tensor concomitant $F[{\bar {\Z}}, {\bar {\sf \Pi}}_{\bar {\Z}}, {\Gamma}]$ reproduces the Poisson bracket action of ${\sf R}({\sf U})[{\sf \Gamma}]$ on $F[{\sf \Gamma}]$, where $F[{\sf \Gamma}]$ is simply $F[{\bar {\Z}}, {\bar {\sf \Pi}}_{\bar {\Z}}, {\Gamma}]$ re-expressed via the canonical transformation (\ref{eq:transf1a})-(\ref{eq:transf4}). Therefore, for two vector fields ${\sf U}_{(1)}$ and ${\sf U}_{(2)}$ on ${\cal U}$, (\ref{eq:aaalgebra}) is again satisfied, and the Lie algebra ${\rm diff}{\yman}$ is antihomomorphically represented by the Poisson algebra of the functionals ${\sf R}{({\sf U})}[{\bar {\Z}}, {\bar {\sf \Pi}}_{\bar {\Z}}]$ on ${\cal G}$.

Our next task is to represent the Lie algebras ${\rm diff}{\tman}$ and ${\rm diff}{\xman}$. This is accomplished by the mapping
\begin{eqnarray}
{R} \; \; : \; \; {\rm diff}{\tman} \times {\rm diff}{\xman} \; \; &\rightarrow& \; \;  {\cal F}[{\cal G}]  \; \; \; \; {\rm by}  
\nonumber
\\
( \, V \in {\rm diff}{\tman}   \, , \,  W \in {\rm diff}{\xman} \, ) \; \; &\mapsto& \; \; 
{R}(V,W)[{\Z},{\PZ}] \in {\cal F}[{\cal G}]
\label{eq:aarmap}
\end{eqnarray} 
where
\begin{equation}
{R}(V,W)[{\Z},{\PZ}] = - {\int}_{\yman} d^2y \, \Big( \, {\Pt} \, V^0({\T}) \, + \,  {\Px} \, W^1({\X})    \Big)(y) \; .
\label{eq:aarmap'}
\end{equation} 
The Poisson brackets of ${R}(V,W)[{\Z},{\PZ}]$ with the history variables identically vanish except
\begin{eqnarray}
\Big\{ \!\!\! \Big\{ \, {\Tt}(y)    \, , \,  {R}(V,W)  \, \Big\} \!\!\! \Big\} &=& - V^0({\T}(y)) \; ,
\nonumber
\\
\Big\{ \!\!\! \Big\{ \, {\Xx}(y)    \, , \,  {R}(V,W)  \, \Big\} \!\!\! \Big\} &=& - W^1({\X}(y)) \; ,
\label{eq:aaI}
\end{eqnarray}
and
\begin{eqnarray}
\Big\{ \!\!\! \Big\{ \, {\Pt}(y)    \, , \,  {R}(V,W) \, \Big\} \!\!\! \Big\} &=&  {\Pt}(y) \, {V^0}_{,0}({\T}(y)) \; ,
\nonumber
\\
\Big\{ \!\!\! \Big\{ \, {\Px}(y)    \, , \,  {R}(V,W)  \, \Big\} \!\!\! \Big\} &=&  {\Px}(y) \, {W^1}_{,1}({\X}(y)) \; .
\label{eq:aaII}
\end{eqnarray} 
Equations (\ref{eq:aaI}) dutifully reproduce what we know about the action of $V$ and $W$ on the maps $\T$ and $\X$, (\ref{eq:V on TX})-(\ref{eq:W on TX}), and on the remaining field variables, (\ref{eq:VW on G}), (\ref{eq:V on gNM})-(\ref{eq:W on gNM}), (\ref{eq:V on fp})-(\ref{eq:W on fp}). Therefore, they also reproduce the action (\ref{eq:V on Y})-(\ref{eq:W on Y}) of $V$ and $W$ on the inverse map (\ref{eq:foliation}):
\begin{eqnarray}
{\cal L}_{({V})}Y^{\alpha}(t,x) =  V^0(t) \, {Y^{\alpha}}_{,0}(t,x)  = \Big\{ \!\!\! \Big\{ \, Y^{\alpha}(t,x)[{\Z}]   \, , \, {R}(V,0)  \, \Big\} \!\!\! \Big\} \; ,
\nonumber
\\
{\cal L}_{({W})}Y^{\alpha}(t,x) = W^1(x) \, {Y^{\alpha}}_{,1}(t,x)  = \Big\{ \!\!\! \Big\{ \, Y^{\alpha}(t,x)[{\Z}]   \, , \,  {R}(0,W)  \, \Big\} \!\!\! \Big\} \; .
\label{eq:aaIII}
\end{eqnarray}

When we calculate the history Poisson bracket between ${R}{\Big( V_{(1)} , W_{(1)} \Big)}$ and ${R}{\Big( V_{(2)} , W_{(2)} \Big)}$, we find that 
\begin{equation}
\Big\{ \!\!\! \Big\{ \, {R}{\Big( V_{(1)} , W_{(1)} \Big)}   \, , \, {R}{\Big( V_{(2)} , W_{(2)} \Big)}  \, \Big\} \!\!\! \Big\} = {R}{\Big( [ \, V_{(1)} \, , \, V_{(2)}  \, ]   ,  [ \, W_{(1)} \, , \, W_{(2)} \,  ]  \Big)} \; .
\label{eq:aaIV}
\end{equation}
We thus see that the map $R$ given by (\ref{eq:aarmap'}) is again an antihomomorphism from the Lie algebras ${\rm diff}{\tman}$ and ${\rm diff}{\xman}$ into the Poisson algebra of the functionals ${\cal F}[{\sf \Gamma}]$ on $\cal G$.

In terms of the new canonical variables ${\bar{\Z}}$, ${\bar {\sf \Pi}}_{\bar {\Z}}$ and ${\Gamma}$ in $\cal G$, the generator (\ref{eq:aarmap'}) takes the form 
\begin{eqnarray}
\!\!\!\!\!\!\!\!\!\!\!\! {R}(V,W)[{\bar {\Z}}, {\bar {\sf \Pi}}_{\bar {\Z}}, {\Gamma}] &=& - {\int}_{\yman} d^2y \, \Big( \, {\bar {\sf \Pi}}_0  \, V^0({\bar {\T}}) \, + \,  {\bar {\sf \Pi}}_1 \, W^1({\bar {\X}}) \Big)(y) 
\nonumber
\\
\!\!\!\!\!\!\!\!\!\!\!\! &+& {\int}_{\zman} d^2z \, \Big( {\pa} \, {\cal L}_{({V})} {\fa}  + {\pab} \, {\cal L}_{({V})} {\gab}  \Big)(t,x) 
\nonumber
\\
\!\!\!\!\!\!\!\!\!\!\!\! &+& {\int}_{\zman} d^2z \, \Big( {\pa} \, {\cal L}_{({W})} {\fa}  + {\pab} \, {\cal L}_{({W})} {\gab}  \Big)(t,x) \; .
\label{eq:aarmap2'}
\end{eqnarray} 
Again, we find that the Poisson bracket action of ${R}(V,W)[{\bar {\Z}}, {\bar {\sf \Pi}}_{\bar {\Z}}, {\Gamma}]$ on an arbitrary tensor concomitant $F[{\bar {\Z}}, {\bar {\sf \Pi}}_{\bar {\Z}}, {\Gamma}]$ reproduces the Poisson bracket action of ${R}(V,W)[{\Z}, {\PZ}]$ on $F[{\sf \Gamma}]$, where $F[{\sf \Gamma}]$ is $F[{\bar {\Z}}, {\bar {\sf \Pi}}_{\bar {\Z}}, {\Gamma}]$ re-expressed via the canonical transformation (\ref{eq:transf1a})-(\ref{eq:transf4}).

In section 10, we obtained the identities (\ref{eq:final Gauss thm}), (\ref{eq:V on S}) and (\ref{eq:U on Sz}), (\ref{eq:VW on S}) expressing the diffeomorphism invariance of the canonical history actions (\ref{eq:as a functional}) and (\ref{eq:in the form}). These identities were derived from the formulae of section 9 which exhibited the action of the vector fields ${\sf U}$, $V$ and $W$ on the arguments ${\sf \Gamma}$ of (\ref{eq:as a functional}) and $\Gamma$ of (\ref{eq:in the form}). We have now learned that all these formulae are correctly reproduced by ${\sf R}(\sf U)$ and $R(V,W)$ in either of their forms (\ref{eq:aarep}), (\ref{eq:aarep2}) and (\ref{eq:aarmap'}), (\ref{eq:aarmap2'}). We are thus able to rewrite the identities (\ref{eq:final Gauss thm}), (\ref{eq:V on S}) and (\ref{eq:U on Sz}), (\ref{eq:VW on S}) in terms of the history Poisson brackets:
\begin{eqnarray}
\Big\{ \!\!\! \Big\{ \, S[{\sf \Gamma}]    \, , \, {\sf R}({\sf U})   \,  \Big\} \!\!\!  \Big\} \equiv 0 \equiv
\Big\{ \!\!\!  \Big\{ \, S[{\sf \Gamma}]     \, , \, R(V,W)   \,  \Big\} \!\!\!  \Big\} \; ,
\label{eq:final equation1}
\\
\Big\{ \!\!\!  \Big\{ \, S[{\Gamma}]     \, , \, {\sf R}({\sf U})   \,  \Big\} \!\!\!  \Big\} \equiv 0 \equiv
\Big\{ \!\!\!  \Big\{ \, S[{\Gamma}]     \, , \, R(V,W)   \,  \Big\} \!\!\!  \Big\} \; .
\label{eq:final equation2}
\end{eqnarray}
Again, the identities (\ref{eq:final equation1}) and (\ref{eq:final equation2}) hold for all vector fields with compact support.  They tell us that the action functionals are invariant under infinitesimal canonical transformations generated by the functionals ${\sf R}(\sf U)$ and $R(V,W)$ which represent infinitesimal diffeomorphisms ${\sf U} \in {\rm diff}{\yman}$, ${V} \in {\rm diff}{\tman}$ and ${W} \in {\rm diff}{\xman}$. This is a fitting conclusion of our endeavor to understand the dynamics of a covariant field system through the history phase space.

\section{Acknowledgments}

We thank P. H\'{a}j\'{\i}\v{c}ek for his comments on our paper. I. K. also appreciates several discussions with K. Savvidou and P. H\'{a}j\'{\i}\v{c}ek about this project. Our research has been supported by the NSF grant PHY9734871 to the University of Utah and by the Tomala Foundation.

\pagebreak

\bibliographystyle{unsrt}

\end{document}